\pdfoutput=1
\documentclass[cernpreprint, UKenglish, texlive=2023, LANGEDIT=true, orcidlogo, texmf, orcidlogo]{atlasdoc}
 
\usepackage[subfig]{atlaspackage}
 
\usepackage{atlasbiblatex}
 
\usepackage{atlasphysics}

\graphicspath{{logos/}{figuresFinal/}}
 
\usepackage{tabularx}
\usepackage{multirow}
\usepackage{comment}
\usepackage{authblk}
\usepackage{hyperref}

\usepackage{ANA-TOPQ-2017-17-PAPER-defs}

\AtlasTitle{Measurement of the top-quark mass using a leptonic invariant mass in $pp$ collisions at $\sqrt{s}=13~\textrm{TeV}$ with the ATLAS detector}

\AtlasAbstract{A measurement of the top-quark mass ($m_t$) in the
$t\bar{t}\rightarrow~\textrm{lepton}+\textrm{jets}$ channel is
presented,  with an experimental technique which exploits semileptonic
decays  of $b$-hadrons produced  in the top-quark decay chain.   The
distribution of the invariant mass $m_{\ell\mu}$ of the lepton,
$\ell$ (with $\ell=e,\mu$), from the $W$-boson decay and the muon,
$\mu$, originating from the $b$-hadron decay is reconstructed, and
a binned-template profile likelihood fit is performed to extract
$m_t$.  The measurement is based on data corresponding to an integrated luminosity of
36.1~fb$^{-1}$ of $\sqrt{s} = 13~\textrm{TeV}$ $pp$ collisions provided by
the Large Hadron Collider and recorded by the ATLAS
detector.  The measured value of the top-quark mass is $m_{t} =
174.41\pm0.39~(\textrm{stat.})\pm0.66~(\textrm{syst.})\pm0.25~(\textrm{recoil})~\textrm{GeV}$,
where the third uncertainty arises from changing the \textsc{Pythia8}
parton shower gluon-recoil scheme, used in top-quark decays, to a recently developed
setup.
}
\author{The ATLAS Collaboration}
 
\AtlasRefCode{ANA-TOPQ-2017-17}
 
\PreprintIdNumber{CERN-EP-2020-018}

\arXivId{2209.00583}

\AtlasJournal{JHEP}
\AtlasJournalRef{JHEP 06 (2023) 019}
\AtlasDOI{10.1007/JHEP06(2023)019}

\hypersetup{pdftitle={ATLAS document},pdfauthor={The ATLAS Collaboration}}
 
\begin{document}
 
\maketitle
 
\tableofcontents

\section{Introduction}
\label{sec:introduction}

The large mass of the top-quark plays a role in much of the dynamics of
elementary particles via loop diagrams. In the Standard Model (SM), the large top-quark mass significantly affects the
radiative corrections to both the Higgs boson and $W$-boson masses, providing a relationship that can be
used for precision tests of the consistency of the SM~\cite{PDG:2018}.
Furthermore, a precise measurement of the top-quark mass is required
to predict the evolution of the Higgs quartic coupling at high
scales~\cite{Degrassi2012, Alekhin:2012py}.
If performed with a precision of the order of a few hundred \MeV, the direct determination of the top-quark mass from its decay products,
and the indirect measurements from top-quark production cross-sections or kinematic distributions,
are important not only for the constraints mentioned above, 
but also for the challenge of the interpretation of such measurements in the context of a strongly interacting particle theory~\cite{Nason:2017cxd, Corcella:arXiv:1903.06574}.
 
A direct measurement of the top-quark mass ($m_t$) is
presented that uses a partial, leptonic-only, invariant mass reconstruction of the top-quark decay products.
The analysis is performed from a sample of reconstructed \ttbar events in the
$\ell+$jets channel, where one of the two $W$-bosons from the top- and antitop-quarks decays leptonically. In the top-quark decay $t\rightarrow Wb$, the invariant mass \mlmu{}
of the lepton $\ell$ (with $\ell=e,\mu$) from the $W$-boson decay and the muon
$\mu$ from a semileptonic decay of a $b$-hadron is constructed as the
observable sensitive to the parent $m_t$ value.
The advantages of a strategy based on the invariant mass of
visible leptonic decay products for the measurement of the top-quark mass rest mainly on the smaller
sensitivity to the jet energy calibration and energy resolution, compared to the
standard direct reconstruction methods, and on less sensitivity to
top-quark production modelling (owing to the boost-invariant
construction) than in methods based on the $W$-decay lepton
alone~\cite{TOPQ-2015-02}. Moreover, methods with different types of systematic uncertainties are
important when combining measurements, and for testing the consistency of the theoretical interpretation of
the top-quark mass.
 
In this analysis the \mlmu{} distribution from models with different top-quark mass hypotheses
is compared with data, and the optimal value of $m_{t}$ is determined from a binned-template profile likelihood fit.
A similar technique was first employed by
the CDF Collaboration at the Tevatron collider~\cite{Aaltonen:2009zi},
and a closely related analysis with \JPsi\ decays has been presented by
the CMS Collaboration~\cite{CMS-TOP-15-014}; however, both these analyses
yielded uncertainties in $m_t$ of several \GeV. Until now, the most precise measurement of the
top-quark mass in the $\ttbar\rightarrow~\ell+$jets channel by the ATLAS
Collaboration was $m_t = 172.08\pm 0.39~\textrm{(stat.)}\pm
0.82~\textrm{(syst.)}$~\GeV, whereas combining multiple ATLAS
measurements gave $m_t = 172.69 \pm 0.48$~\GeV~\cite{TOPQ-2017-03}.
The CMS Collaboration reports its most precise combination as $m_t = 172.44 \pm 0.49$~\GeV~\cite{CMS-TOP-14-022}, and the Tevatron
experiments report a combined value of $m_t = 174.30 \pm 0.65$~\GeV~\cite{TEVEWWG}.
Finally, the mass of the top-quark is indirectly determined from global
electroweak fits as $m_t = 176.4\pm 2.1$~\GeV~\cite{EWfits}.


\section{ATLAS experiment}
\label{sec:detector}

The ATLAS experiment~\cite{PERF-2007-01} at the LHC is a multipurpose particle detector with a forward--backward symmetric cylindrical geometry and a near 4$\pi$ coverage in solid angle\footnote{ATLAS uses a right-handed coordinate system with its origin at the
nominal interaction point (IP) in the centre of the detector and the $z$-axis along the beam pipe. The $x$-axis points from the IP to the centre of
the LHC ring, and the $y$-axis points upwards. Cylindrical coordinates
($r$,$\phi$) are used in the transverse plane, $\phi$ being the azimuthal angle
around the $z$-axis. The pseudorapidity is defined in terms of the
polar angle $\theta$ as $\eta=-\ln\tan(\theta/2)$. Angular
distance is measured in units of $\Delta R\equiv\sqrt{(\Delta\eta)^2+(\Delta\phi)^2}$.}.
It consists of an inner
tracking detector surrounded by a thin superconducting solenoid
providing a 2~T axial magnetic field, electromagnetic and hadronic
calorimeters, and a muon spectrometer. The inner tracking detector
covers the pseudorapidity range $|\eta| < 2.5$ and consists of silicon
pixel, silicon microstrip, and transition radiation tracking detectors.
The innermost layer, known as the insertable B-layer~\cite{ATLAS-TDR-2010-19,PIX-2018-001},
was added in 2014 and provides high-resolution hits at small
radius to improve the tracking performance.
Lead/liquid-argon (LAr) sampling calorimeters provide electromagnetic
(EM) energy measurements with high granularity.
A steel/scintillator-tile hadronic calorimeter covers the central
pseudorapidity range ($|\eta| < 1.7$). The endcap and forward regions
are instrumented with LAr calorimeters for both the EM and hadronic
energy measurements up to $|\eta| = 4.9$. The muon spectrometer
surrounds the calorimeters and is based on three large air-core toroid
superconducting magnets with eight coils each and a bending power of
2.0 to 7.5~Tm. It includes a system of precision tracking chambers covering the region $|\eta| < 2.7$
and fast detectors for triggering in the range $|\eta| < 2.4$.
A two-level trigger system was used to select events~\cite{TRIG-2016-01}.
The first-level trigger is implemented in hardware and uses a subset of the detector information to reduce the accepted rate to at most 100~kHz.
This is followed by the software-based high-level trigger, which reduces the event rate to around 1~kHz.
An extensive software suite~\cite{ATL-SOFT-PUB-2021-001} is used in the reconstruction and analysis of real and simulated data, in detector operations, and in the trigger and data acquisition systems of the experiment.


\section{Data and Simulation}
\label{sec:data_and_simulation}

\subsection{Data sample and object definition}
\label{sec:data_object_definition}
The analysis is performed with the 2015 and 2016 proton--proton
collision data sample produced by the LHC at a centre-of-mass energy of $\sqrt{s}=13~\TeV$ and
collected by the ATLAS experiment, corresponding to an integrated luminosity of
36.1~fb$^{-1}$~\cite{ATLAS-CONF-2019-021}.
The data sample was recorded during stable beam conditions, and all relevant ATLAS detector subsystems were required to be operational.
The average number of $pp$
collisions in the same bunch crossing (referred to as pile-up) in the data sample is 24.1.
 
Electron candidates are reconstructed from energy deposits (clusters) in the electromagnetic calorimeter
matched to reconstructed tracks in the inner detector. Candidates in the
transition region $1.37 <|\eta_\textrm{cluster}| < 1.52$ between the calorimeter barrel and endcaps are
excluded. Muon candidates are reconstructed from track segments in the layers of the muon spectrometer,
and matched to tracks found in the inner detector.
The final muon candidates are re-fitted using the complete track information from both detector systems.
Jet candidates are reconstructed from three-dimensional topological EM-scale energy
clusters~\cite{PERF-2014-07} in the calorimeter using the \antikt\ jet
algorithm~\cite{Cacciari:2008gp,Fastjet} with a radius parameter $R=0.4$. The
reconstructed jets are calibrated to the level of
stable-particle jets by the application of a jet energy scale (JES) correction derived from simulation and
\textit{in situ} corrections based on $13~\TeV$ data~\cite{PERF-2016-04}. There is no dedicated energy scale correction for jets with semileptonic heavy-flavour hadron decays.
The missing transverse momentum, $\met$, is defined as the magnitude of the
negative vector sum of the transverse momentum, \pt, of all reconstructed and calibrated physics objects in the
event, with an extra term added to account for soft energy in the event that is
not associated with any of the reconstructed objects~\cite{PERF-2016-07}. This soft term is calculated from
inner-detector tracks matched to the primary vertex in order to make it more
resilient to pile-up contamination.
\subsection{Object and event selections}
\label{sec:object_event_selection}
The event selection is designed to collect a sample of \ttbar candidate events in the final
state $\ell\nu b jj^\prime\bar{b}$, where $\ell=e$ or $\mu$ and the
$jj^\prime$ are the jets produced in the decay of the $W$-boson into quarks,
and at least one of the $b$-initiated jets is associated with a muon from the semileptonic
decay of a $b$-hadron. The goal is to select events where the lepton $\ell$ from the $W$-boson
and the $b$-initiated jet with the muon from
semileptonic decay come from the same top-quark.
 
Events are required to pass either a single-electron or single-muon
trigger. Multiple trigger types were used: the lowest-threshold triggers include isolation requirements
to reduce the trigger rate and had
$p_\textrm{T}$ thresholds of $20~\GeV$ for muons and $24~\GeV$ for electrons in
2015 data, and $26~\GeV$ for both lepton types in 2016 data~\cite{TRIG-2018-05, TRIG-2016-01, TRIG-2018-01}.
These triggers were complemented by others with higher $\pt$
thresholds and no isolation requirements to increase event
acceptance. Events must have at least one reconstructed vertex,
i.e.\ at least two tracks with $\pt > 0.4~\GeV$ consistent with the
beam-collision region in the $x$--$y$ plane. If multiple vertices
are reconstructed, then the primary vertex is taken to be the one with the largest sum,
over  the tracks assigned to it, of the  transverse momentum squared of each track.
 
Events are further selected based on the presence of an
electron or muon candidate from the decay of a $W$-boson, called `primary' leptons.
Primary-lepton electron candidates must satisfy a `tight'
likelihood-based identification criterion~\cite{PERF-2017-01}, be matched to the
corresponding trigger, and have $\pt>27~\GeV$, $|\eta|<2.47$ with the
exclusion of $1.37<|\eta|<1.52$,
longitudinal impact parameter $|z_0 \sin \theta| < 0.5$~mm and transverse impact parameter significance
$|d_0 / \sigma(d_0)| < 5$, where $\sigma(d_0)$ is the
uncertainty in the transverse impact parameter.
Background from photon conversions, hadrons, and electrons produced away from the primary
vertex (`non-prompt' electrons) is reduced by requiring the primary electron candidates to pass an
isolation requirement based on the surrounding tracks and topological
clusters in the calorimeter~\cite{PERF-2017-01}.
Primary-lepton muon candidates must satisfy a `medium'
quality identification criterion~\cite{PERF-2015-10}, be matched to the
corresponding trigger, and have $\pt>27$~\GeV, $|\eta|<2.5$, longitudinal impact parameter $|z_0 \sin \theta| < 0.5$~mm and transverse impact parameter significance
$|d_0 / \sigma(d_0)| < 3$. Background from hadrons and from muons produced
away from the primary vertex (`non-prompt' muons) is reduced by requiring primary muon candidates
to pass an isolation requirement based on the surrounding tracks and topological
clusters in the calorimeter, and be separated by $\Delta R>0.4$ from
the nearest selected jet. If the nearest selected jet is
$\Delta R\leq0.4$ from the muon and has less than three associated tracks (including the muon track), the muon is kept
and the jet is removed from the jet list, to ensure high efficiency for muons undergoing significant energy loss in the calorimeter.
Events with more than one candidate primary lepton with $\pt>25~\GeV$
are vetoed, in order to reject events from the \ttbar dileptonic decay channel.
 
Jet candidates are required to have $\pt>25~\GeV$ and $|\eta|<2.5$, and
a multivariate jet-vertex-tagger (JVT) is applied to suppress jets from pile-up~\cite{PERF-2014-03}.
During jet reconstruction, no distinction is made between
identified electrons and jet energy deposits. Therefore, if any of the jets lie
within $\Delta R=0.2$ of a selected electron, the single closest jet is
discarded in order to avoid double-counting electrons as jets. After this,
electrons that are within $\Delta R=0.4$ of a remaining jet are
removed. Jets are identified as originating from a $b$-quark ($b$-tagged) using
two techniques, one based on the reconstruction of a displaced jet (DJ tagging) and the other based
on the semileptonic decay of a $b$-hadron into a so-called `soft
muon' (SMT tagging). For DJ tagging,
multivariate techniques are used to combine
information about the impact parameters of displaced tracks and the
topological properties of secondary and tertiary decay vertices reconstructed
within the jet~\cite{PERF-2016-05}. The algorithm is trained on simulated \ttbar\ events to discriminate
$b$-jets from a background consisting of light-flavour jets and
$c$-jets. A selection corresponding to an efficiency of $77\%$ for
$b$-jets in \ttbar\ events is employed, with a rejection rate of a factor of 7 (100) for $c$-jets (light jets). The SMT tagging is performed
by requiring the presence of a muon candidate satisfying the `tight' quality identification criterion~\cite{PERF-2015-10},
$\pt>8~\GeV$ and $|\eta|<2.5$, with loose requirements on the impact
parameters ($|d_0|<3$~mm, $|z_0 \sin \theta|<3$~mm) and with a
distance $\Delta R<0.4$ from a selected jet candidate.
The definition of the muon object for the SMT tagging
was optimised by maximising the efficiency for muons
originating from the semileptonic decays of $b$- and $c$-hadrons (selecting approximately 50\% of $b$-jets
containing a muon, which are in turn 20\% of all $b$-jets
produced in \ttbar\ events), minimising the misidentification rate
(about $10^{-3}$ per light-flavour jet, mostly due to the decays of pions and kaons), and minimising the uncertainty in the
measured top-quark mass.
If more than one muon satisfying the criteria above is found within a given jet,
the muon with the highest \pt\ is chosen.  
 
Events must have at least one SMT-tagged jet and at least one DJ-tagged jet (which
could be the same jet), among a total of at least four jet candidates with $\pt>30~\GeV$ (with the exception of the SMT-tagged
jet which may have a $\pt$ as low as 25~\GeV). If more than one SMT-tagged jet is found in the event, only the one with the highest-\pt{} muon is considered. The SMT muon and the
primary lepton must be separated by $\Delta R_{\ell,\mu}<2$.
The presence of at least one neutrino in the final state is inferred from the requirements
that $\met>30~\GeV$ and $\met + \mtw>60~\GeV$.\footnote{The
transverse mass is given by
$\mtw = \sqrt{2 p^\ell_{\mathrm T} \met (1-\cos\Delta\phi)}$, where
$p^\ell_{\mathrm T}$  is the transverse momentum of the
muon (electron) and $\Delta\phi$ is the azimuthal angle separation
between the lepton and the direction of the missing transverse
momentum.}
The requirement that the SMT muon and the
primary lepton must be separated by $\Delta R_{\ell,\mu}<2$ enhances the fraction of events where both leptons come from the same
top-quark, in contrast to events where the two leptons originate from different top-quarks.
The selected events are categorised as same-sign (SS)
events or opposite-sign (OS) events according to the charge signs of the primary lepton
and the soft muon. When both leptons come from the same top-quark, opposite-sign events are enriched in direct
$b\rightarrow\mu X$ decays, while same-sign events have a large contribution from
sequential $b\rightarrow c X^\prime \rightarrow \mu X^{''}$ decays; both
samples carry information about the mass of the parent top-quark though.
Finally, the invariant mass of the primary lepton and the soft
muon (\mlmu{}) is required to be between 15 and 80~\GeV, as this is the
region most sensitive to the top-quark mass. This requirement also
suppresses the $Z$-boson, $J/\psi$ and $\Upsilon$ resonances.

\subsection{Signal and background simulations}
\label{sec:mc_simulations}
A number of Monte Carlo (MC) simulation samples are used to model the expected
signal of top-quark pairs and the background.
The MC samples were processed either through the full ATLAS detector simulation~\cite{SOFT-2010-01} based
on {\scshape Geant4}~\cite{Agostinelli:2002hh} or through a faster simulation making use of parameterised
showers in the calorimeters~\cite{ATL-PHYS-PUB-2010-013}.
Additional simulated $pp$ collisions generated by {\scshape Pythia}~8.186~\cite{Sjostrand:2007gs} with the
\textsc{MSTW2008}~\cite{mstwnnlo1,mstwnnlo2} set of leading-order (LO) parton distribution functions (PDFs)
and a set of tuned parameters called the A2 tune~\cite{ATL-PHYS-PUB-2012-003} were used to model
the effects of both in-time and out-of-time pile-up.
They were superimposed on the MC events, matching the luminosity profile of the recorded data.
All simulated samples were processed through the same reconstruction
algorithms and analysis chain as the data.
 
Simulated MC events were
corrected so that the object identification efficiencies and energy and
momentum scales and resolutions matched those determined from data control
samples~\cite{PERF-2017-03,PERF-2015-10,PERF-2016-04,PERF-2014-03}.
The modelling of SMT muons and their misidentification due to light-hadron decays and detector background (`SMT fake') was studied
using control samples as well.  The efficiency of muon identification in
jets was calibrated using muons from \JPsi\ and $Z$ decays, and
checked as a function of nearby track and calorimeter
activity, and of the muon's transverse impact parameter $d_0$.
The calibration of the misidentification rate was performed, using the sample of $W$+jets
events described in the following of the section, with the same approach used in Ref.~\cite{TOPQ-2016-07}.
A data-driven technique, described in detail in Ref.~\cite{TOPQ-2012-08}, is used to measure in data the normalisation and
flavour composition of such events and to derive corrections that are applied to simulated samples for the calibration. A
light-jets dominated sample of $W$+ 1 jet events is then defined by selecting events where the jet is SMT-tagged but is not
DJ-tagged, and using the SS category. A data-to-simulation scale factor (SF) of $1.10\pm0.14$ is measured. A slight miscalibration of the \pt\ of jets that contain a soft muon
was observed, and the \pt\ of these jets was corrected in the
simulation with a factor of $0.967\pm0.024$. It has been measured by
studying the distribution of the ratio of the \pt\ values of the SMT-tagged jet and the
average non-SMT-tagged jet in \ttbar data and simulation.

The \ttbar sample was generated using the $hvq$ program~\cite{Frixione:2007nw}
in the \textsc{Powheg-Box}~v2  generator~\cite{Nason:2004rx,Alioli:2010xd} with the
\NNPDF[3.0nlo] set of PDFs~\cite{Ball:2014uwa}
and the top-quark mass set to 172.5~\GeV. Additional samples with different top-quark mass
hypotheses were produced in the range of $m_{t}$ between 165 and
180~\GeV, with steps of $0.5~\GeV$ between 170 and 175~\GeV. The samples have been produced using the appropriate top-quark decay width values predicted at next-to-next-to-leading order
(NNLO) as a function of $m_{t}$~\cite{PhysRevLett.110.042001}.
The $hvq$ program uses on-shell matrix elements for production of \ttbar\ pairs at next-to-leading order (NLO) in
quantum chromodynamics (QCD).
Off-shell effects and top-quark decays, including spin correlations,
were approximated using
\textsc{Madspin}~\cite{Frixione:2007zp}.  Parton showers and
hadronisation were modelled by \textsc{Pythia}~8.2~\cite{Sjostrand:2007gs}
using a dedicated `A14-$r_b$' setting of the ATLAS
A14~\cite{ATL-PHYS-PUB-2014-021}  tune, as detailed in
Section~\ref{sec:hq-fragmentation}. The A14 tune is based on LEP and
Tevatron collider data and also uses a combination of ATLAS $\sqrt{s}$=7 \TeV{} measurements of the underlying event, jet production, $Z$-boson production and
top-quark production in order to constrain the parameters for the
showers, multiple parton interactions and colour reconnection effects.
Radiation in top-quark decays was  handled entirely by the parton-shower generator, which implements
matrix-element corrections with an accuracy equivalent to a calculation at the
NLO level. The $h_\textrm{damp}$ parameter, which controls the \pt\ of the first additional
emission beyond the Born configuration, was set to 1.5 times the top-quark mass
of each sample. The main effect of the $h_\textrm{damp}$ setting is to regulate the
high-\pt\ emission against which the \ttbar\ system recoils.
The \textsc{EvtGen}~v1.2.0~\cite{Lange:2001uf} program was used to
simulate bottom and charm hadron mixing and decays. The production fractions and the branching ratios (BR) of the
decay of $b$-hadrons and $c$-hadrons into muons were rescaled to the latest
values from the Particle Data Group (PDG)~\cite{PDG:2018}, as detailed
in Section~\ref{sec:hq-fragmentation}.
The simulated \ttbar\ event sample was normalised to the Top++2.0~\cite{Czakon:2011xx} theoretical cross-section of $832^{+46}_{-51}$~pb, calculated at NNLO in QCD and including resummation of next-to-next-to-leading logarithmic
(NNLL) soft gluon terms~\cite{ref:xs1,ref:xs2,ref:xs3,ref:xs4,ref:xs5}.
 
The main backgrounds to candidate signal events come from the
production of a single top-quark, and from a $W$- or $Z$-boson in
association with jets. A small background contribution arises from diboson ($WW$, $WZ$, $ZZ$) production.
Events not containing prompt leptons also contribute to the
selected sample via the misidentification of a jet or a photon as an
electron, or the presence of non-prompt electrons or muons passing the prompt isolated lepton selection.
This contribution is referred to as `multijet' background, and was
estimated using data by following the matrix method described in Ref.~\cite{ATLAS-CONF-2014-058}.
 
The production of $V+$jets was simulated with the
\SHERPA[2.2.1]~\cite{Bothmann:2019yzt}
generator using next-to-leading-order (NLO) matrix elements (ME) for up to two partons, and leading-order (LO) matrix elements
for up to four partons calculated with the Comix~\cite{Gleisberg:2008fv}
and \OPENLOOPS~\cite{Buccioni:2019sur,Cascioli:2011va,Denner:2016kdg} libraries. They
were matched with the \SHERPA parton shower~\cite{Schumann:2007mg} using the \MEPSatNLO
prescription~\cite{Hoeche:2011fd,Hoeche:2012yf,Catani:2001cc,Hoeche:2009rj}
with the set of tuned parameters developed by the \SHERPA authors.
The \NNPDF[3.0nnlo] set of PDFs~\cite{Ball:2014uwa} was used and the samples
were normalised to a next-to-next-to-leading-order (NNLO)
prediction~\cite{Anastasiou:2003ds}.
The normalisation of the $W$+jet background and the fractions
of $W$-bosons produced in association with heavy-flavour quarks are extracted from data, taking
advantage of the intrinsic $W$-boson charge asymmetry in this process~\cite{TOPQ-2017-01}. The
$Z$+jets contribution is estimated from MC simulation and checked in
a data control sample.
 
Diboson processes were simulated with the \SHERPA[2.1.1] event generator. They were calculated using Comix and \OPENLOOPS{}, and merged with the \SHERPA parton shower according to the \MEPSatNLO prescription.
The \CT[10nlo] PDF set~\cite{Lai:2010vv} was used in conjunction with dedicated parton shower tuning developed by the \SHERPA authors. 
 
Samples of $t-$, $Wt$- and $s$-channel single top-quark background events were generated with
\textsc{Powheg-Box}~v1, using the 4-flavour scheme for the NLO matrix element calculations together with the fixed four-flavour PDF set \CT[10f4].
Overlaps between the \ttbar\ and $Wt$ final states were
removed with the `diagram removal' prescription~\cite{Frixione:2008yi}.
For all top processes, top-quark spin correlations were preserved (for t-channel, top quarks were decayed using \textsc{Madspin}).
All single top-quark samples were interfaced to \textsc{Pythia}~6.428~\cite{Sjostrand:2006za} with the CTEQ6L1 PDF set~\cite{Pumplin:2002vw} and the corresponding
Perugia~2012~\cite{Skands:2010ak} set of tuned parameters. The \textsc{EvtGen}~v1.2.0 program was used to model
properties of the bottom and charm hadron decays.
The single top-quark $t$- and $s$-channel samples were normalised to the
approximate NNLO theoretical cross-sections~\cite{Kidonakis:2010ux,Kidonakis:2010tc,Kidonakis:2011wy}.
\subsection{Modelling of heavy-quark fragmentation, hadron production and decays}
\label{sec:hq-fragmentation}
The modelling of the momentum transfer between
the $b$-quark and the $b$-hadron is an important aspect of this
analysis. The Monte Carlo event generators, such as the \textsc{Pythia}, \textsc{Herwig}~\cite{Bahr:2008pv,Bellm:2015jjp} and \SHERPA programs,
describe this transfer according to phenomenological models, namely
the string and cluster models containing parameters which are tuned to
data. The \textsc{Pythia8} program uses parametric functions to describe the $b$-quark fragmentation function,
while \textsc{Herwig7} and \SHERPA use a non-parametric model which handles the complete parton-shower evolution.
The free parameters in those models are typically fit to
measurements from $e^+e^-$ colliders, and this analysis assumes that the
$b$-quark fragmentation function is the same in $e^+e^-$ and $pp$ collisions, as supported by dedicated studies.
 
The Lund--Bowler parameterisation~\cite{Andersson:1983ia,Bowler} in  \textsc{Pythia8} was used. It  is given by
 
\begin{equation*}
\label{eq:LB}
f(z) = \frac{1}{z^{1+b r_b m^2_b}}(1-z)^{a}\exp(-b m^2_\textrm{T}/z),
\end{equation*}
 
where $a$, $b$ and $r_b$ are the function parameters, $m_b$ is the $b$-quark mass,
$m_\textrm{T}=\sqrt{m_{B}^2+\pt^2}$ the $b$-hadron transverse mass ($m_{B}$ being the $b$-hadron mass), and $z$ is the fraction of the longitudinal energy carried by the $b$-hadron with respect to the $b$-quark, in the light-cone
reference frame.
The fragmentation function is defined at the hadronisation scale
and it is evolved by the parton shower to the process scale through DGLAP evolution equations.
In \textsc{Pythia8}, the values of $a$ and $b$ were fit
to data sensitive to light-quark fragmentation~\cite{Skands:2014pea},
such as charged-particle multiplicities, event shapes and scaled momentum distributions.
They are assumed to be universal for light- and heavy-quarks,
while the $r_b$ parameter is specific to $b$-quark fragmentation.
 
\begin{figure}
\centering
\includegraphics[width=0.7\textwidth]{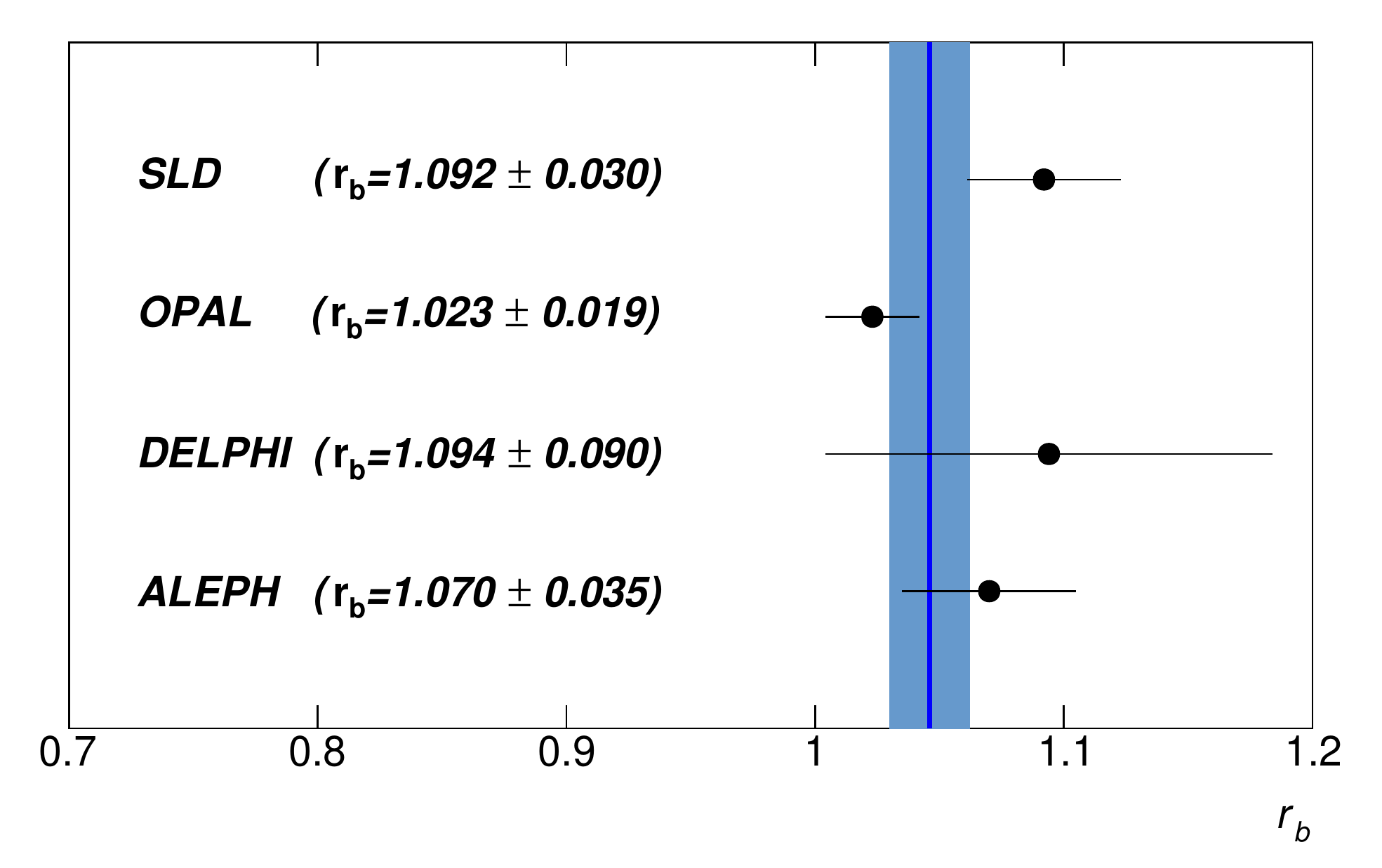}
\caption{Fit of the $r_b$ \textsc{Pythia} parameter to the four experiments considered: SLD, OPAL, DELPHI and ALEPH. For each experiment, the $r_b$ fitted value is shown together with its uncertainty. These results have been obtained including bin-by-bin correlations in the
experimental data (except for SLD experiment) and after rescaling the uncertainties on $r_b$ for DELPHI and SLD experiments as explained in the text. The solid blue line represents the value of $r_b$ obtained from the
combined fit to the four experiments. The blue band represents the uncertainty from the combined fit to the four experiments.}
\label{fig:rb_LEP_fit_corr}
\end{figure}

The description of the $b$-quark fragmentation in the ATLAS A14
tune is improved by fitting for the \texttt{StringZ:rFactB} \textsc{Pythia8}
parameter (corresponding to $r_b$) following the approach given in
Refs.~\cite{Corcella:2005,Corcella:2010,Corcella:2010b,Corcella:2018}. The A14 tune sets the parton shower $\alpha_s$ to
0.127, whereas the value of 0.1365 is used in Monash~\cite{Skands:2014pea}. However, both Monash and A14 set $r_b=0.855$.
Since the $b$-quark fragmentation is controlled both by $\alpha_s$ and $r_b$, the procedure described in the following is
used to determine a value of $r_b$ more appropriate for a value of $\alpha_s = 0.127$.
The fit uses the A14 tune with $e^+e^-$
collision data from the ALEPH, DELPHI and OPAL experiments at the LEP collider, and from the
SLD experiment at the SLC collider~\cite{ALEPH4,DELPHI6,OPAL5,SLD}.
The distribution of $x_B = 2p_B \cdot
p_Z /m^2_Z$ from semileptonically decaying $b$-hadrons in
$e^+e^-\rightarrow Z\rightarrow b\bar{b}$
events is used, where $p_B$ and $p_Z$ are the four-momenta of the
$b$-hadron and the $Z$-boson, respectively.
In the $Z$ rest frame,  $m_Z$ is twice the beam energy and therefore
$x_B = 2E_B / m_Z$, where $E_B$ is the energy of the $b$-hadron.
The fit is performed using \textsc{Rivet}~v3.1.0~\cite{Buckley:2019ar} to implement the
measurements. 
The effect of the matrix-element corrections for $e^+e^-\rightarrow Z\rightarrow b\bar{b}g$
is taken into account. Eighty simulated samples of 1M $e^+e^-\rightarrow Z\rightarrow b\bar{b}$ events were produced using \textsc{Pythia8}
with different values of the $r_b$ parameter in the interval [0.8-1.4] and compared to the experimental data in HEPDATA format. The extraction of the best $r_b$ value is performed through a standard binned $\chi^2$ test on the experimental $x_B$ distribution where statistical and systematic
uncertainties are taken into account for each of the four
experiments. In addition, for the results of ALEPH, DELPHI and OPAL,
bin-by-bin correlations are taken into account in the fit
procedure. The SLD experiment did not provide the full covariance
matrix for the total uncertainties and therefore the $\chi^2$ fit for this experiment is performed ignoring bin-by-bin correlations. For each experiment, the $\chi^2$ minimisation is performed and the best $r_b$ value and its uncertainty are found. The results are summarised in
Table~\ref{tab:rb_corr}. The values of $\chi^2/\textrm{ndf}$ for DELPHI and SLD experiments show a poor modelling of the data by the simulated templates.
Therefore, before including these results into a global $\chi^2$ combination, the
uncertainties found in the $r_b$ parameter for these two cases were rescaled by a factor of $S=\sqrt{\chi^2/\textrm{ndf}}$ following the procedure outlined in Ref.~\cite{PDG:2018}. After these uncertainties are rescaled, the four $\chi^2$ curves
are summed up to produce a single $\chi^2$ curve taking into account the information of all four experiments. In this approach, the four experiments are considered uncorrelated since the dominant uncertainties on the measurements come from uncorrelated sources.
This curve is then fitted with a parabola to find the minimum of the
$\chi^2$ and to extract the best value of the $r_b$ parameter and its
uncertainty, yielding $r_b=1.05\pm0.02$. The values of the $r_b$ parameter and its uncertainty for the four experiments, and their combination are shown in Figure~\ref{fig:rb_LEP_fit_corr}.
A similar procedure is performed, as a cross-check, exploiting the
average value <$x_B$> of the four $x_B$ distributions. The
experimental <$x_B$> values are compared with the predicted value of the various simulated
samples, and the best $r_b$ values and their uncertainties are
extracted. A weighted average of the single results is calculated and
the result is found to be compatible with the previous method.
 
\begin{figure}[h]
\centering
\subfloat[]{\includegraphics[width=0.45\textwidth]{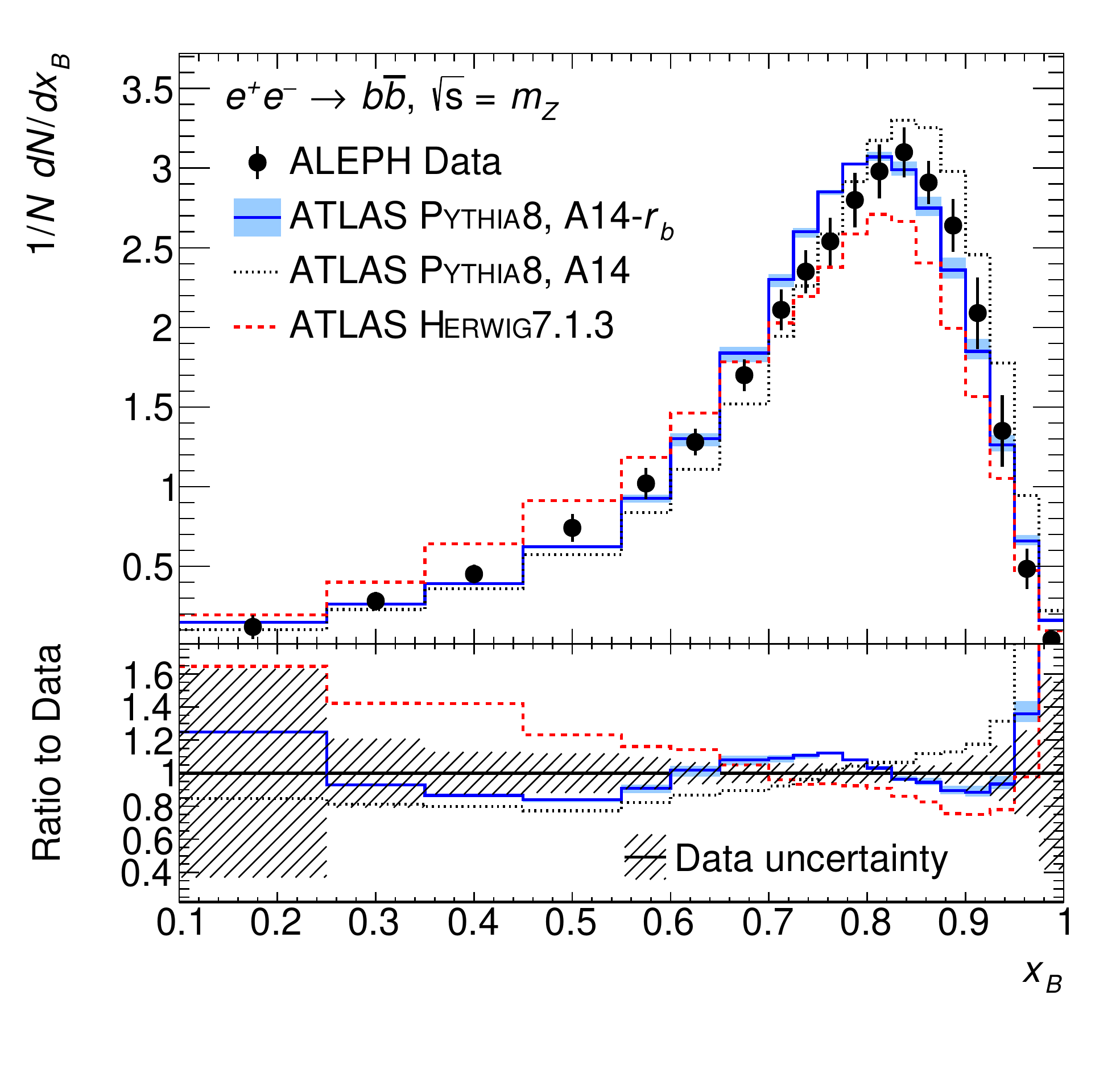}}
\subfloat[]{\includegraphics[width=0.45\textwidth]{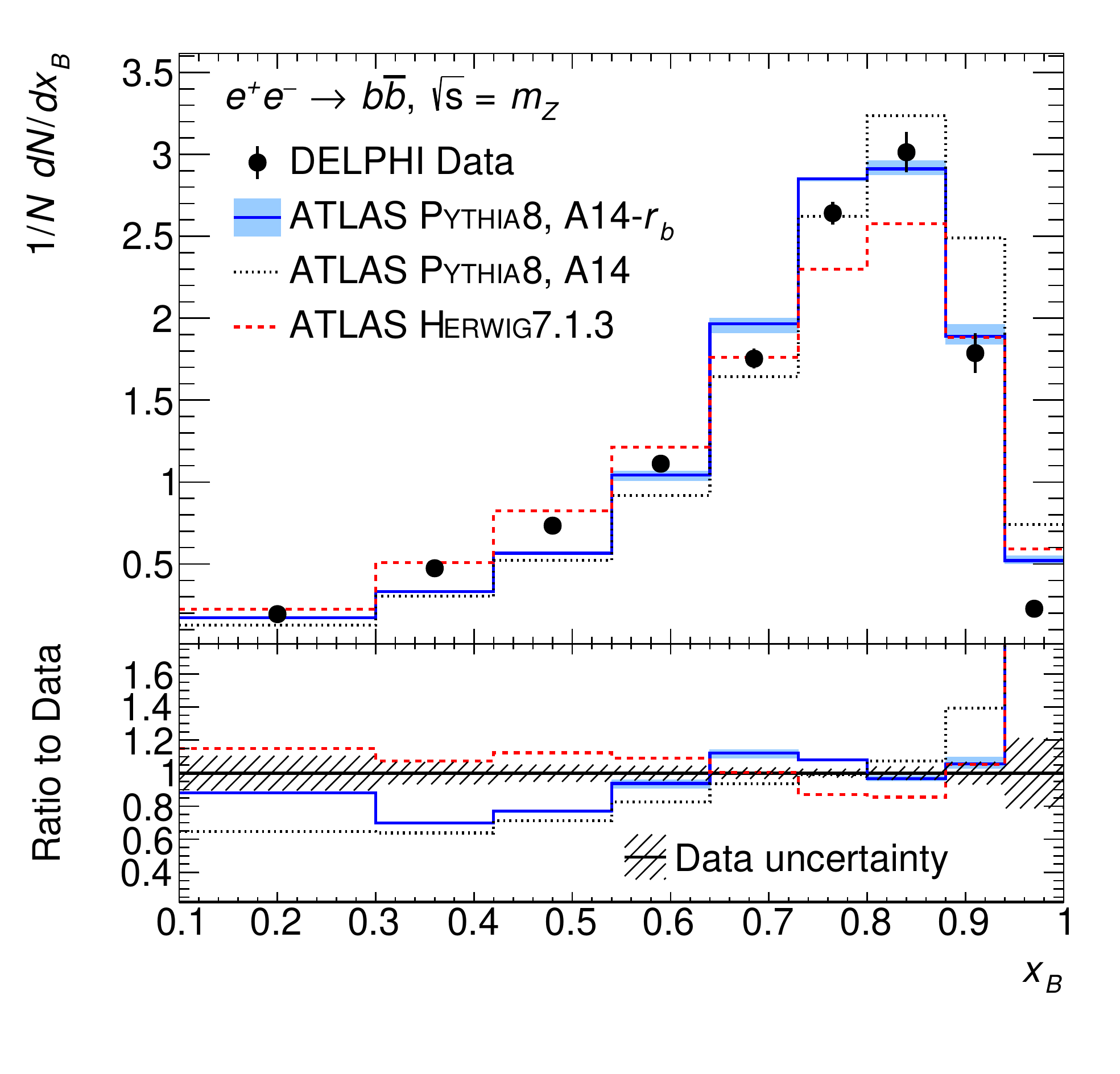}}\\
\subfloat[]{\includegraphics[width=0.45\textwidth]{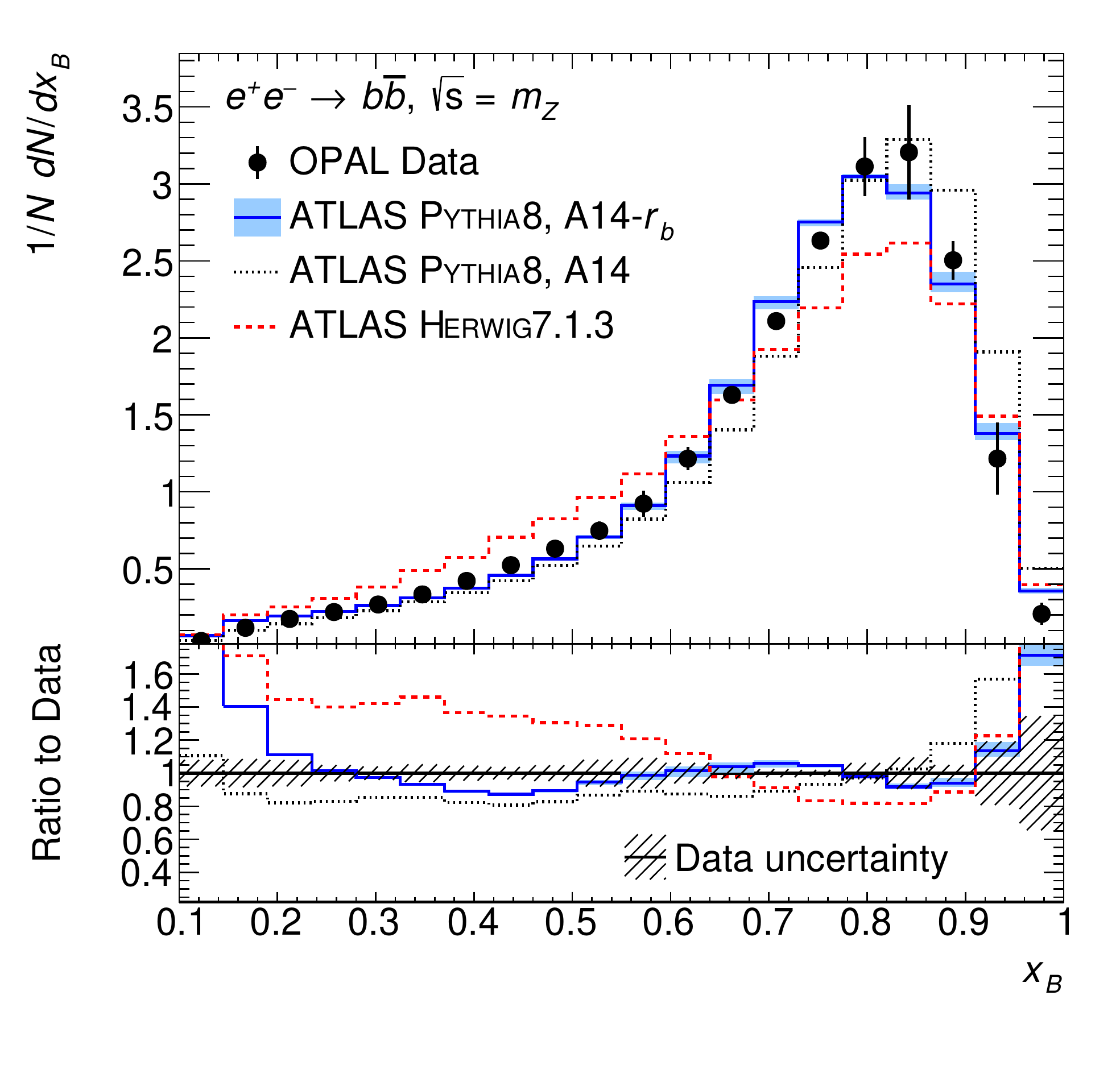}}
\subfloat[]{\includegraphics[width=0.45\textwidth]{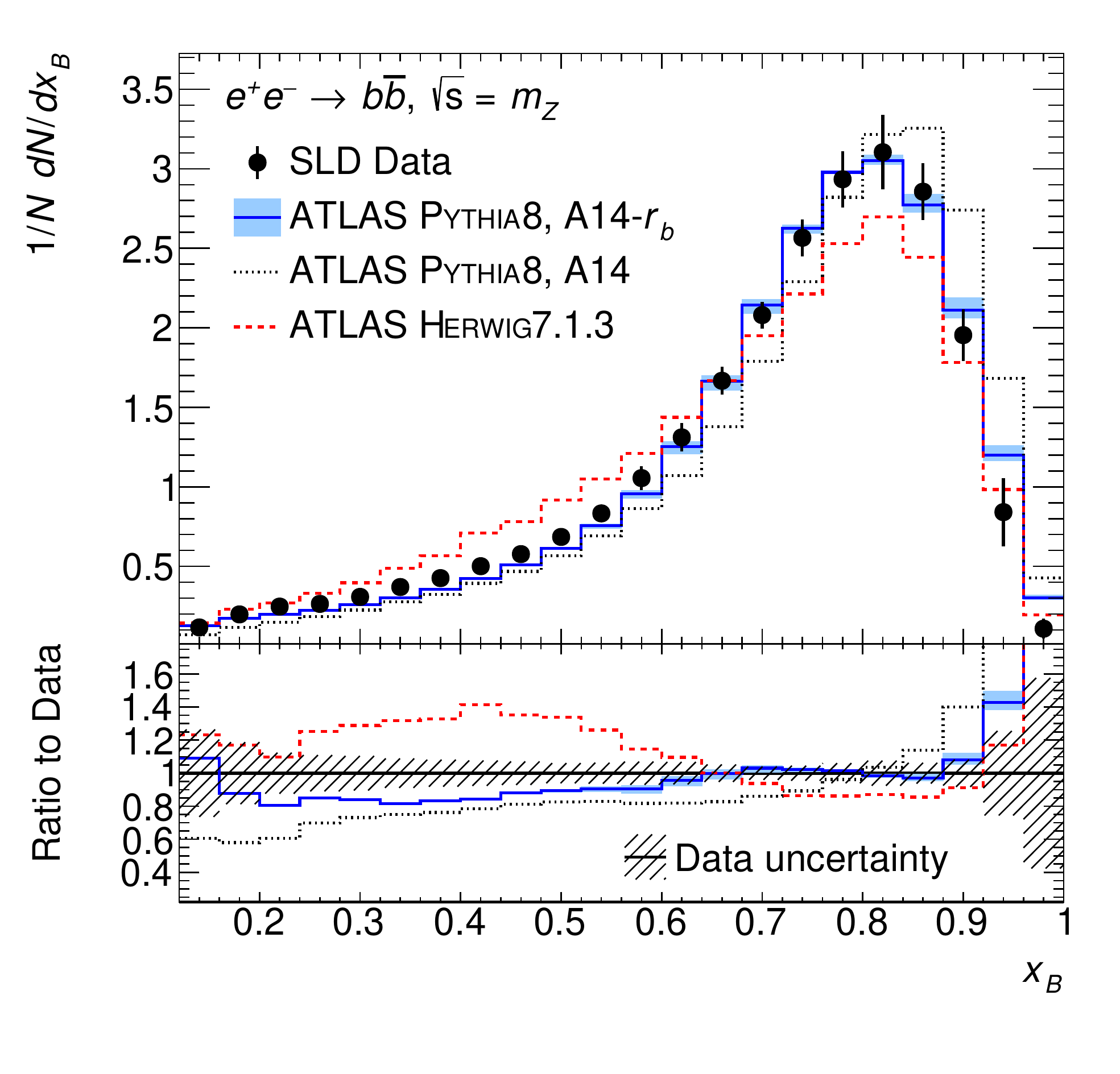}}
\caption{Distributions of $x_B$ from semileptonically decaying $b$-hadrons in $e^+e^-\rightarrow Z\rightarrow b\bar{b}$ events as measured by the ALEPH, DELPHI and OPAL experiments at the LEP collider, and by the SLD experiment at the SLC collider.
Each of the measured spectra is compared with a set of MC predictions. The \textsc{Pythia8} A14-$r_b$ prediction includes a systematic uncertainty band obtained considering the $r_b$ parameter variations described in the text and included as systematic
uncertainty in the analysis.}
\label{fig:xB_LEP}
\end{figure}

The setting with $r_b=1.05\pm0.02$ is referred to as A14-$r_b$ and
is applied in this analysis to all signal MC samples
using \textsc{Pythia8} for the simulation of the parton shower. The $x_B$ distributions measured by the ALEPH, DELPHI, OPAL and SLD
experiments are shown in Figure~\ref{fig:xB_LEP} together with the MC predictions obtained with the A14, the A14-$r_b$ and the \textsc{Powheg}+\textsc{Herwig7} settings respectively. Recent $b$-fragmentation measurements based on LHC data~ \cite{STDM-2018-52,TOPQ-2017-19} show reasonable agreement with simulations using the  A14-$r_b$ tune.
 
\begin{table}
\caption{Results of a fit  of the $r_b$ parameter of the Lund-Bowler $b$-quark fragmentation function
in \textsc{Pythia8}  to  different experimental data where bin-by-bin correlations for ALEPH, DELPHI
and OPAL experiments were considered.}
\centering
\begin{tabular}{lcr}
\hline
Experiment & $r_b$ & $\chi^2/\textrm{ndf}$\\
\hline
ALEPH  & $1.070 \pm 0.035 $ & 21/18 \\
DELPHI & $1.094 \pm 0.030 $ & 73/8 \\
OPAL   & $1.023 \pm 0.019 $ & 18/19 \\
SLD    & $1.092 \pm 0.018 $ & 58/21 \\
\hline
\end{tabular}
\label{tab:rb_corr}
\end{table}
 
The production fractions of weakly decaying $b$- and $c$-hadrons observed in
\textsc{Powheg}+\textsc{Pythia8} MC simulation with \textsc{EvtGen} are rescaled to those
from the Heavy Flavour Averaging Group (HFLAV)~\cite{HFAG} as reported
by the PDG~\cite{PDG:2018} and in Ref.~\cite{c_had_prodfrac}. The production fraction values and
corresponding scale factors for \textsc{Powheg}+\textsc{Pythia8} simulations
are shown in Table~\ref{tab:prodfracPyt}. These scale factors refer only to the first weakly decaying hadron produced in the hadronisation process of $b$- and $c$-quarks. The scale factors are applied to each of these hadrons present in a MC simulated
event, with the overall event weight given by the product of these
scale factors. This procedure assumes that the production fractions of
heavy-flavour quarks can be regarded as universal in the kinematic
phase space relevant for this analysis, within the uncertainties
accounted for here, as supported by recent
results~\cite{LHCb:frag1,LHCb:frag2,LHCb:frag3,LHCb:frag4,LHCb:frag5,LHCb:frag6,ALICE:frag1}
which find deviations from universality only in the very low-\pt{} regime.
 
\begin{table}
\centering
\caption{The production fraction values for $b$-hadrons and
$c$-hadrons in the PDG and \textsc{Powheg}+\textsc{Pythia8}. The
relative scale factors applied to \textsc{Powheg}+\textsc{Pythia8} are also shown.
The values in the PDG column are derived from Refs.~\cite{PDG:2018}
and~\cite{c_had_prodfrac}. The same scale factors are applied to the
charge-conjugate hadrons.}
\begin{tabular}{llcc}
\toprule
Hadron   & PDG     & \textsc{Powheg}+\textsc{Pythia8}    & Scale Factor \\ \midrule
$B^0$       & $0.404 \pm 0.006$ & 0.429 & $0.941 \pm 0.014$ \\
$B^+$       & $0.404 \pm 0.006$ & 0.429 & $0.942 \pm 0.014$ \\
$B^0_s$      & $0.103 \pm 0.005$ & 0.095 & $1.088 \pm 0.052$ \\
$b$-baryon & $0.088 \pm 0.012$ & 0.047 & $1.87 \pm 0.26$ \\ \bottomrule
$D^+$       & $0.226 \pm 0.008$ & 0.290 & $0.780 \pm 0.027$ \\
$D^0$       & $0.564 \pm 0.015$ & 0.553 & $1.020 \pm 0.027$ \\
$D^0_s$      & $0.080 \pm 0.005$ & 0.093 & $0.857 \pm 0.054$ \\
$c$-baryon & $0.109 \pm 0.009$ & 0.038 & $2.90 \pm 0.24$ \\ \bottomrule
\end{tabular}
\label{tab:prodfracPyt}
\end{table}
 
The  branching ratios of the $b$- and $c$-hadron decays that contain a
muon are also adjusted to match those measured by previous experiments~\cite{PDG:2018}. Central values and relative
scale factors, along with the corresponding uncertainties, are shown
in Table~\ref{tab:BRs}; these are measured after applying the corrections for the production fractions described above.
The $b$ $\to$ $\bar{c}$  $\to$ $\mu$ branching ratio was determined
by averaging the direct measurement from DELPHI~\cite{DELPHI5} and the
predicted values computed by the LEP Electroweak Heavy
Flavour Working Group~\cite{lephf}. The latter prediction is based on
flavour-specific $B \to D$ and $B \to \Lambda^+_c$ rates measured at
CLEO~\cite{cleo1,cleo2,cleo3} in combination with the $B \to DD(X) $
rates measured in ALEPH~\cite{ALEPH5} to extract the probabilities of
producing the different $c$-hadrons from the initial $b$-hadron
decays.
The $c$-hadron semileptonic branching fractions were also used in the
prediction.
The $c\to\mu$ scale factor is applied only to the semileptonic decays
of $c$-hadrons into muons when the $c$-hadrons do not come from a
cascade $b$-hadron decay.
\begin{table}
\centering
\caption{Hadron to muon branching ratios from the PDG and in \textsc{Powheg}+\textsc{Pythia8}+\textsc{EvtGen}. The relative scale factors applied to \textsc{Powheg}+\textsc{Pythia8}
are also shown. The  values in the PDG column are derived from Refs.~\cite{PDG:2018}
and~\cite{c_had_prodfrac}. The $c\to\mu$ scale factor is applied only to the semileptonic decays
of $c$-hadrons into muons when the $c$-hadrons do not come from a
cascade $b$-hadron decay. The same scale factors are applied
to the charge conjugate hadrons.}
\label{tab:BRs}
\begin{tabular}{l l @{ }l l l @{ }l}
\toprule
Hadronic    &  \multicolumn{2}{c}{PDG}      & \textsc{Powheg}    &    \multicolumn{2}{c}{Scale}  \\
Decay Mode &   \multicolumn{2}{c}{ }           & \textsc{Pythia8}+\textsc{EvtGen}     &\multicolumn{2}{c}{Factor} \\
\midrule
$b$ $\to$ $\mu$                   & 0.1095 & $^{+0.0029}_{-0.0025}$       & 0.106                          & 1.032 & $^{+0.0027}_{-0.0023}$           \\
$b$ $\to$ $\tau$        	  &  0.0042 & $\pm 0.0004$               & 0.0064                         & 0.661& $\pm$0.062           \\
$b$ $\to$ $c$ $\to$ $\mu$         &  0.0802 & $\pm 0.0019$               & 0.085                & 0.946 & $\pm$0.022          \\
$b$ $\to$ $\bar{c}$  $\to$ $\mu$ &  0.016 & $\pm 0.003$        & 0.018                          & 0.89 & $\pm$0.17         \\
$c$ $\to$ $\mu$                   &  0.082 & $\pm 0.005 $                & 0.084                          & 0.976 & $\pm$0.059        \\ \bottomrule
\end{tabular}
\end{table}


\section{Analysis}
\label{sec:analysis}

\subsection{Event yields and sample composition}
The number of observed candidate events and the predicted signal and background are shown in
Table~\ref{tab:yields_bTagAnyOptiOS}, for both the OS and SS regions.
Over 90\% of the events in the sample contain a top-quark pair, including
cases where the soft muon is erroneously chosen from a \ttbar dilepton
decay, whereby a muon from the prompt $W$ decay is found near a jet or
radiates a near-collinear photon mimicking a soft muon tag, and cases where the soft muon candidate does not
originate from a $b$ decay such as in $W\rightarrow cs$ or SMT fakes.
The contributions from single top-quark,
$W$- or $Z$-boson in association with jets, and multijet
background are not negligible. The $Z$+jets background gives a small
contribution near the peak of the \mlmu{} distribution,
but becomes important for \mlmu{} close to the $Z$-boson mass peak.
 
Of the selected $t\bar{t}$ events in the OS class, 83\% are cases where the primary
lepton and the soft muon belong to the decay of the same top-quark,
and 10\% are events where the two originate from different top-quarks.
The topological requirement $\Delta R_{\ell,\mu}<2$ is responsible for much of the purity because
it is very effective in preferentially selecting the same top-quark decays.
For the remaining 7\% of cases,  the soft muon 
does not originate from either of the $b$-quarks produced in the two top-quark decays.
In the SS class, the above fractions are 57\%, 41\% and 2\%, respectively.
The rate of soft muons that are not from $b$-quarks is higher in OS events
owing to the decay of $c$-quarks that come from
$W\rightarrow c s$ in the top-quark decay chain.
 
\sisetup{separate-uncertainty,table-format=6.1(6)}
\begin{table}[thbp]
\begin{center}
\caption{Event yields with \mlmu{} between 15 and 80 GeV, separately
for OS and SS regions. Uncertainties
shown include statistical and systematic contributions but not the
recoil uncertainty.}
\label{tab:yields_bTagAnyOptiOS}
\begin{tabular}{lSS}
\toprule
Process      & \multicolumn{1}{c}{Yield (OS)} &  \multicolumn{1}{c}{Yield (SS)}\\
\midrule
$t\bar{t}$ (SMT from $b$- or $c$-hadron)       & 55700 \pm 3400     & 34800 \pm 2300 \\
$t\bar{t}$ (SMT from $W\rightarrow \mu\nu$)   & 2190 \pm 310            &  4.9 \pm 3.6 \\
$t\bar{t}$ (SMT fake)   & 1490 \pm 210 & 1240 \pm 170 \\
Single top $t$-channel   & 770 \pm 70 & 490 \pm 40 \\
Single top $s$-channel   & 63 \pm 6 & 49 \pm 4 \\
Single top $Wt$ channel & 1840 \pm 140 & 1260 \pm 100 \\
$W$+jets   &  1600 \pm 400 & 1080 \pm 240 \\
$Z$+light jets   & 210 \pm 80 & 15 \pm 6\\
$Z$+HF jets   & 550 \pm 180 & 310 \pm 100 \\
Diboson   & 17.2 \pm 2.9 & 6.3 \pm 1.4 \\
Multijet   & 530 \pm 140 & 480 \pm 130 \\
\hline
Total Expected & 65000 \pm 4000  & 39700 \pm 2500 \\
\hline
Data   & 66891  & 42087\\
\bottomrule
\end{tabular}
\end{center}
\end{table}
 
In order to better understand the nature of the sample composition in the OS and SS
regions, the expected \ttbar\ events shown in the first row of
Table~\ref{tab:yields_bTagAnyOptiOS} are further resolved into components. Table~\ref{tab:fraction_optimOS} shows the components involving direct and sequential decays, and decays not belonging
to the $b$-quark from the $t\rightarrow Wb$ transition. The direct $t \rightarrow  B \rightarrow \mu$ decays are by far the dominant component in the OS sample, while the sequential $t \rightarrow B \rightarrow D \rightarrow \mu$ decays are also suppressed by the kinematic requirement on the soft muon. Decay channels involving $\tau$ leptons give a small contribution.
\begin{table}[htbp]
\sisetup{ table-number-alignment=left,group-digits=false}
\begin{center}
\caption{Fraction of MC-simulated \ttbar\ events with a soft muon originated by a $b$- or $c$-hadron split into
components of direct and sequential decays, and decays not belonging
to the $b$-quark from the $t\rightarrow Wb$ transition, separately for the
opposite-sign and same-sign event selections. The letters $B$
and $D$ indicate $b$- and $c$-hadrons of either charge. Only MC events
with two real muons are included.}
\begin{tabular}{l | S[table-format=2.2] | S[table-format=2.2]}
\toprule
& \multicolumn{1}{c}{OS [\%]} & \multicolumn{1}{c}{SS [\%]}   \\ \hline
Processes involving a $\mu$ from a $t$ or $\bar{t}$ & & \\
$t \rightarrow  B \rightarrow \mu$ & 73.6 & 51.2 \\
$t \rightarrow B \rightarrow D \rightarrow \mu$ & 16.7 & 44.2 \\
$t \rightarrow B \rightarrow \tau \rightarrow \mu$ & 2.0 & 1.3 \\
$t \rightarrow B \rightarrow D \rightarrow \tau \rightarrow \mu$ & 0.8 & 0.8 \\
\midrule
Processes involving a $\mu$ not from a $t$ or $\bar{t}$ & & \\
$B\rightarrow \mu $ & 0.6 & 0.9\\
$D\rightarrow \mu$ & 5.8 & 1.4 \\
Other ($\tau \rightarrow \mu$) & 0.5 & 0.1 \\
\bottomrule
\end{tabular}
\label{tab:fraction_optimOS}
\end{center}
\end{table}
 
\begin{figure}[!h]
\centering
\subfloat[]{\includegraphics[width=0.45\textwidth]{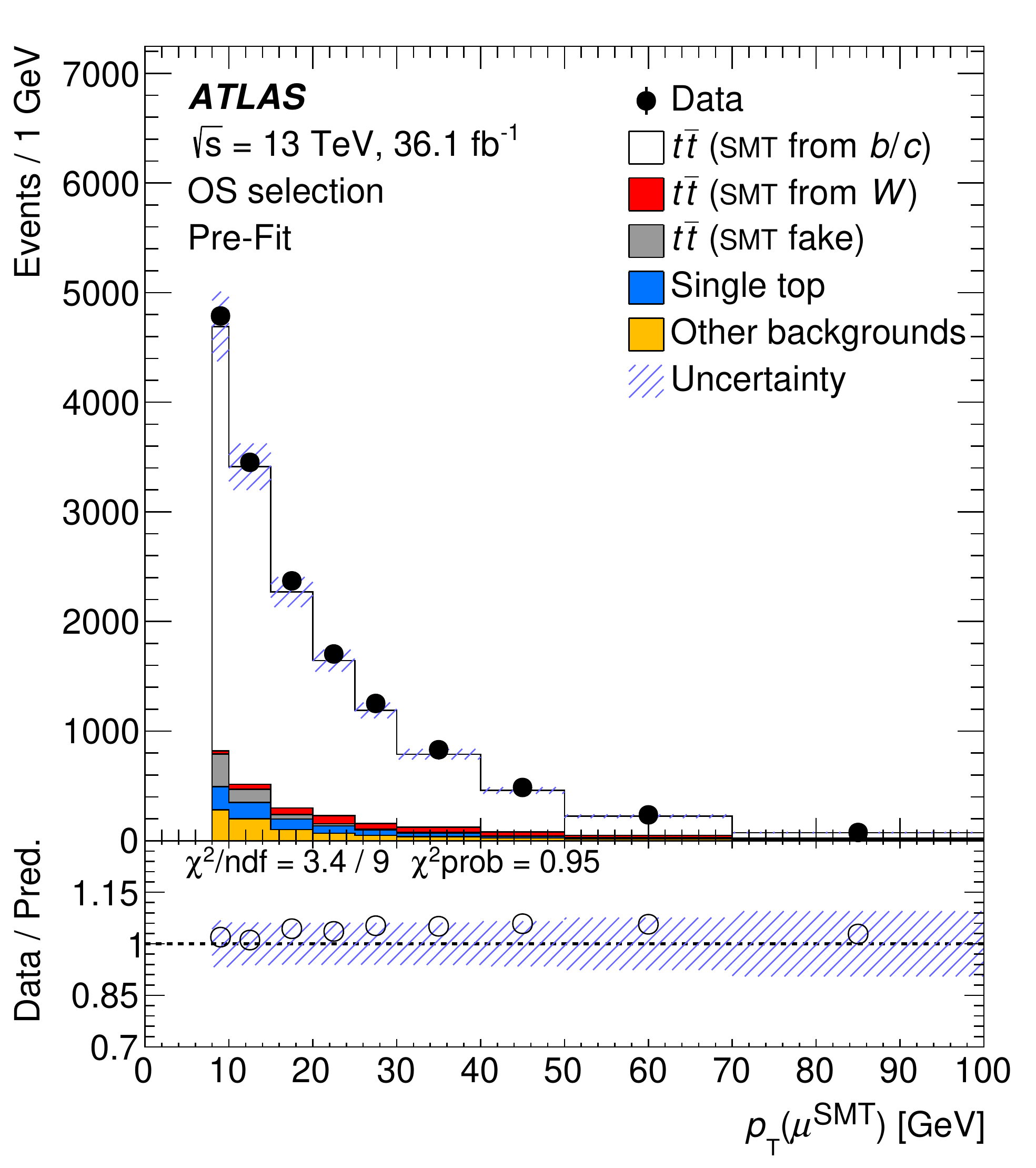}}
\subfloat[]{\includegraphics[width=0.45\textwidth]{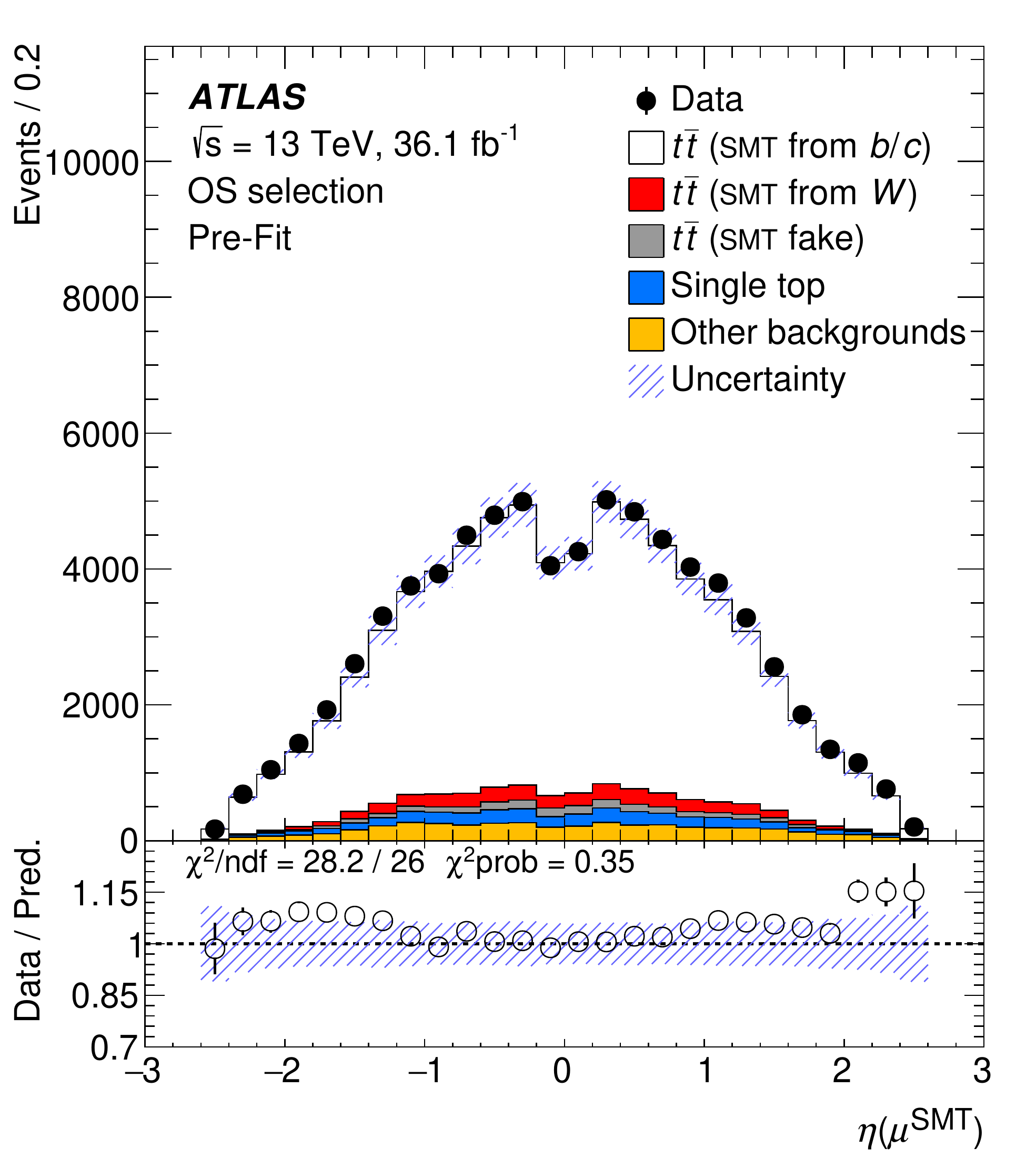}}
\\
\subfloat[]{\includegraphics[width=0.45\textwidth]{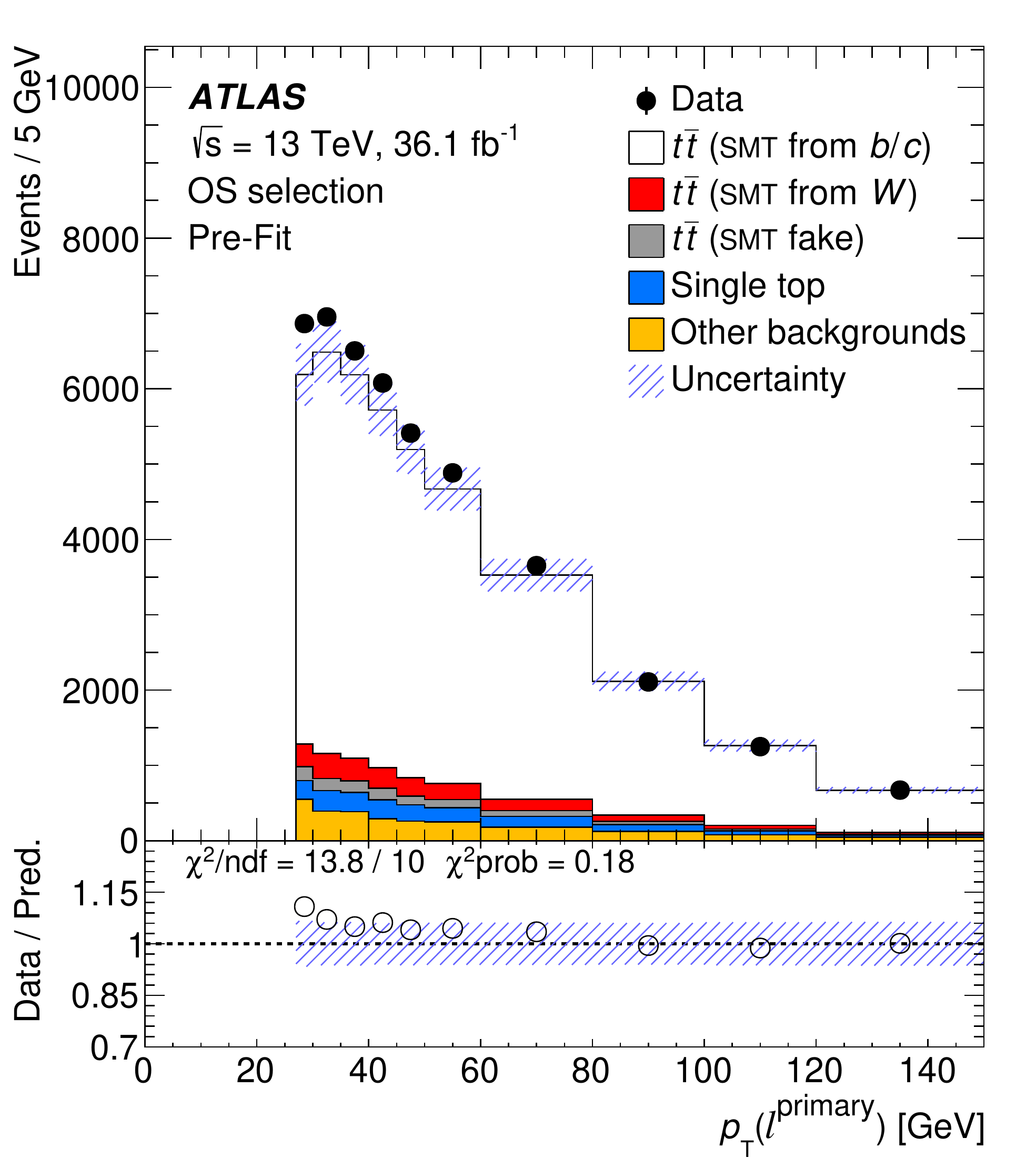}}
\subfloat[]{\includegraphics[width=0.45\textwidth]{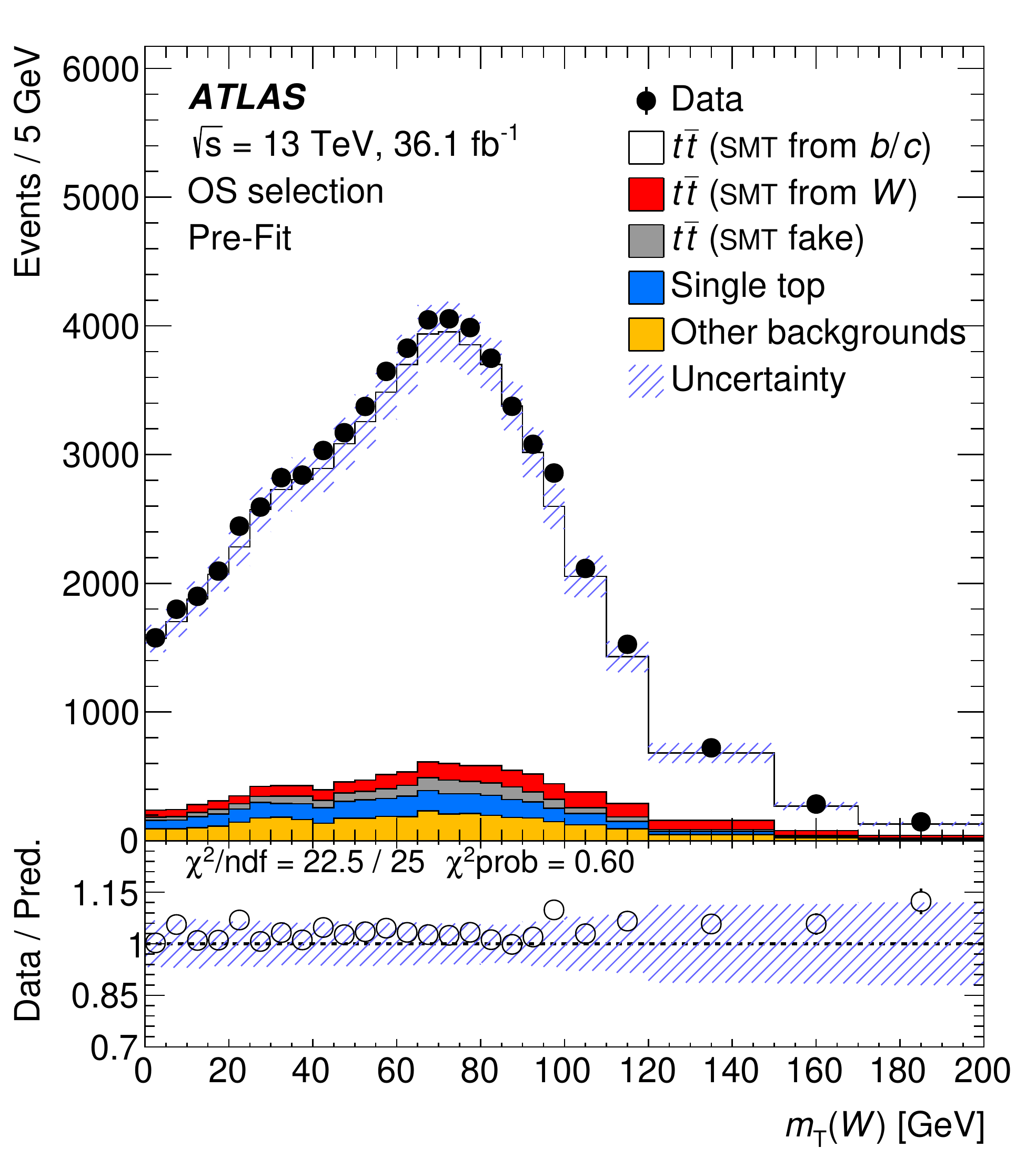}}
\caption{Comparison of data and prediction before the fit
described in Section~\ref{sec:extraction_of_the_top_mass}
in the OS sample, i.e. for events with primary lepton and the soft muon with opposite charges, for the (a) soft muon \pt, (b) soft muon $\eta$,
(c) primary lepton \pt\ and (d) $W$-boson transverse mass. The prediction reports the expected event contribution from the
signal and backgrounds. The  uncertainty band includes statistical
and systematic uncertainties, but does not include the recoil uncertainty.}
\label{fig:plots_bTagAnyOpti_Calib_OS}
\end{figure}

\begin{figure}[!h]
\centering
\subfloat[]{\includegraphics[width=0.45\textwidth]{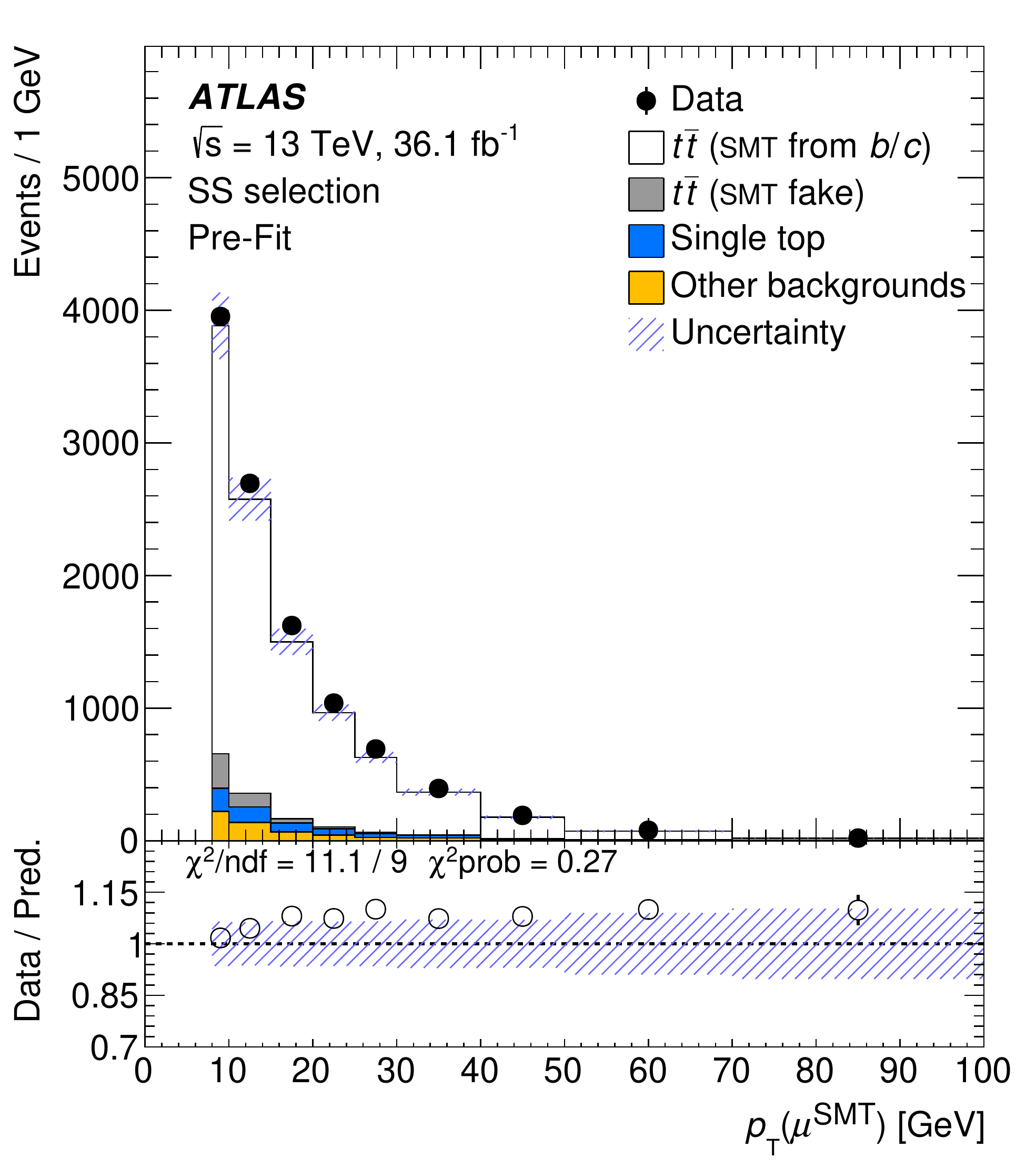}}
\subfloat[]{\includegraphics[width=0.45\textwidth]{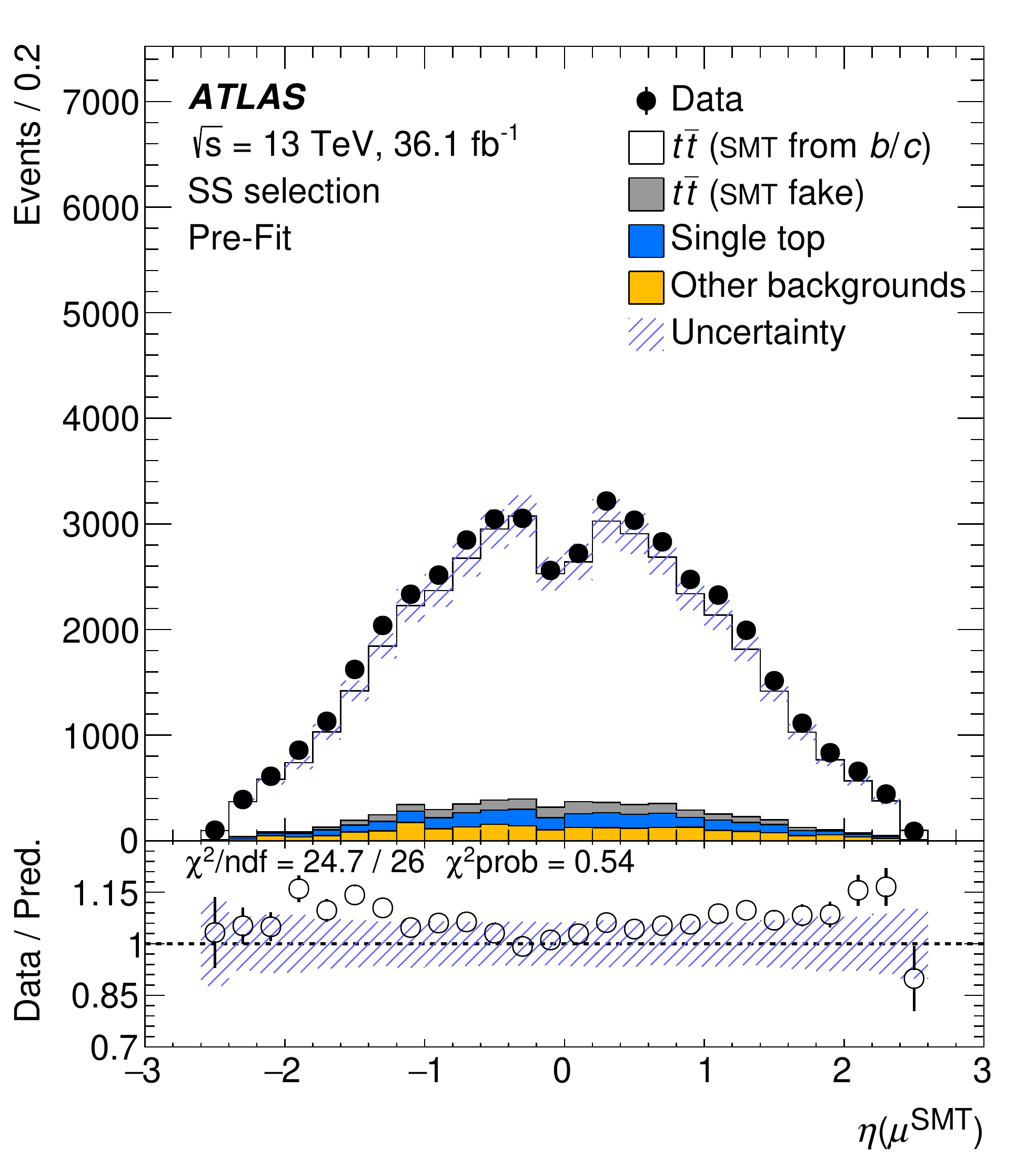}}
\\
\subfloat[]{\includegraphics[width=0.45\textwidth]{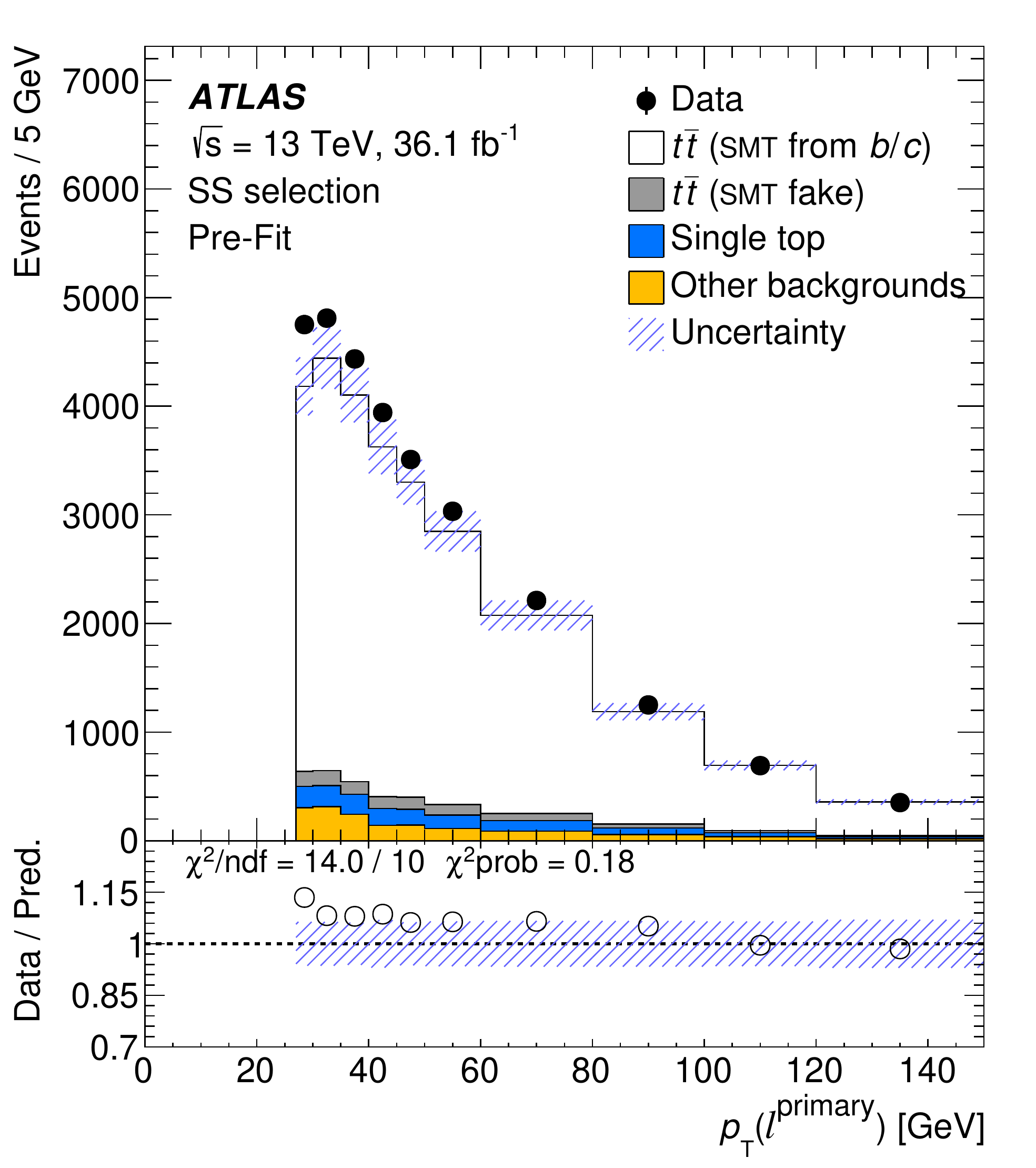}}
\subfloat[]{\includegraphics[width=0.45\textwidth]{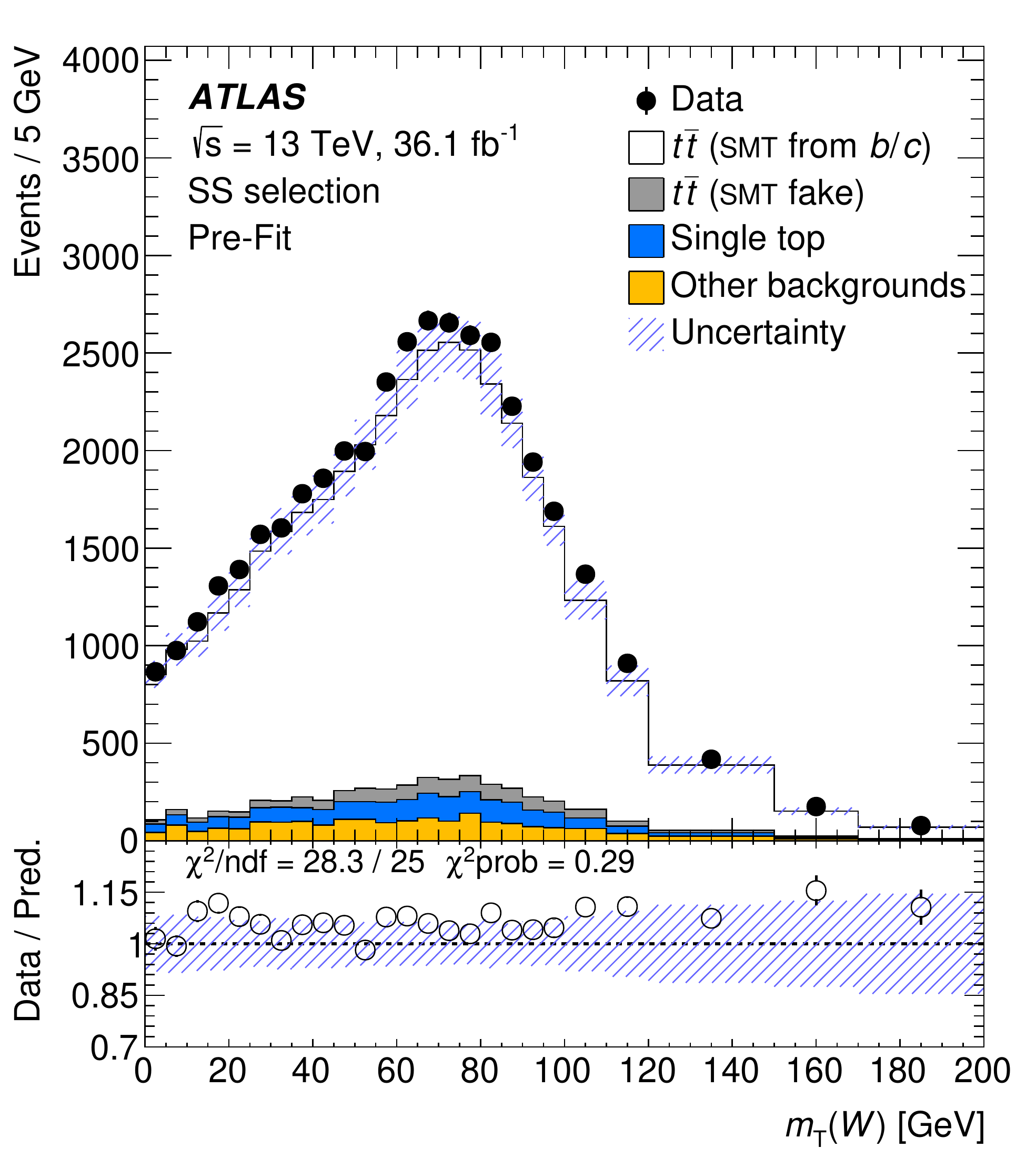}}
\caption{Comparison of data and prediction before the fit
described in Section~\ref{sec:extraction_of_the_top_mass}
in the SS sample, i.e. for events with primary lepton and the soft muon with same charges, for the (a) soft muon \pt, (b) soft muon $\eta$,
(c) primary lepton \pt\ and (d) $W$-boson transverse mass. The prediction reports the expected event contribution from the
signal and backgrounds. The  uncertainty band includes statistical
and systematic uncertainties, but does not include the recoil uncertainty.}
\label{fig:plots_bTagAnyOpti_Calib_SS}
\end{figure}

The data are compared with the sum of the predicted signal and
backgrounds in Figures~\ref{fig:plots_bTagAnyOpti_Calib_OS} and~\ref{fig:plots_bTagAnyOpti_Calib_SS},
for an illustrative selection of kinematic distributions of the candidate
events for the OS and SS selections: $\pt(\mu^\textrm{SMT})$, $\eta(\mu^\textrm{SMT})$,
$\pt(\ell^\textrm{~primary})$ and $\mtw$. The uncertainties are
discussed in detail in Section~\ref{sec:systematic_uncertainties}. The
compatibility of the data and MC predictions is studied using a
$\chi^2$ test involving the bin-by-bin full correlation matrix, and for
all distributions the level of agreement is better than 2
standard deviations. The slight excess of SMT muons in regions of high-$\eta$ is associated with a potential small mismodelling of the efficiency for low-\pt{} muons in simulation.
The corresponding impact on the measurement is negligible.
 
\subsection{Extraction of the top-quark mass}
\label{sec:extraction_of_the_top_mass}
\begin{figure}
\centering
\subfloat[]{\includegraphics[width=0.45\textwidth]{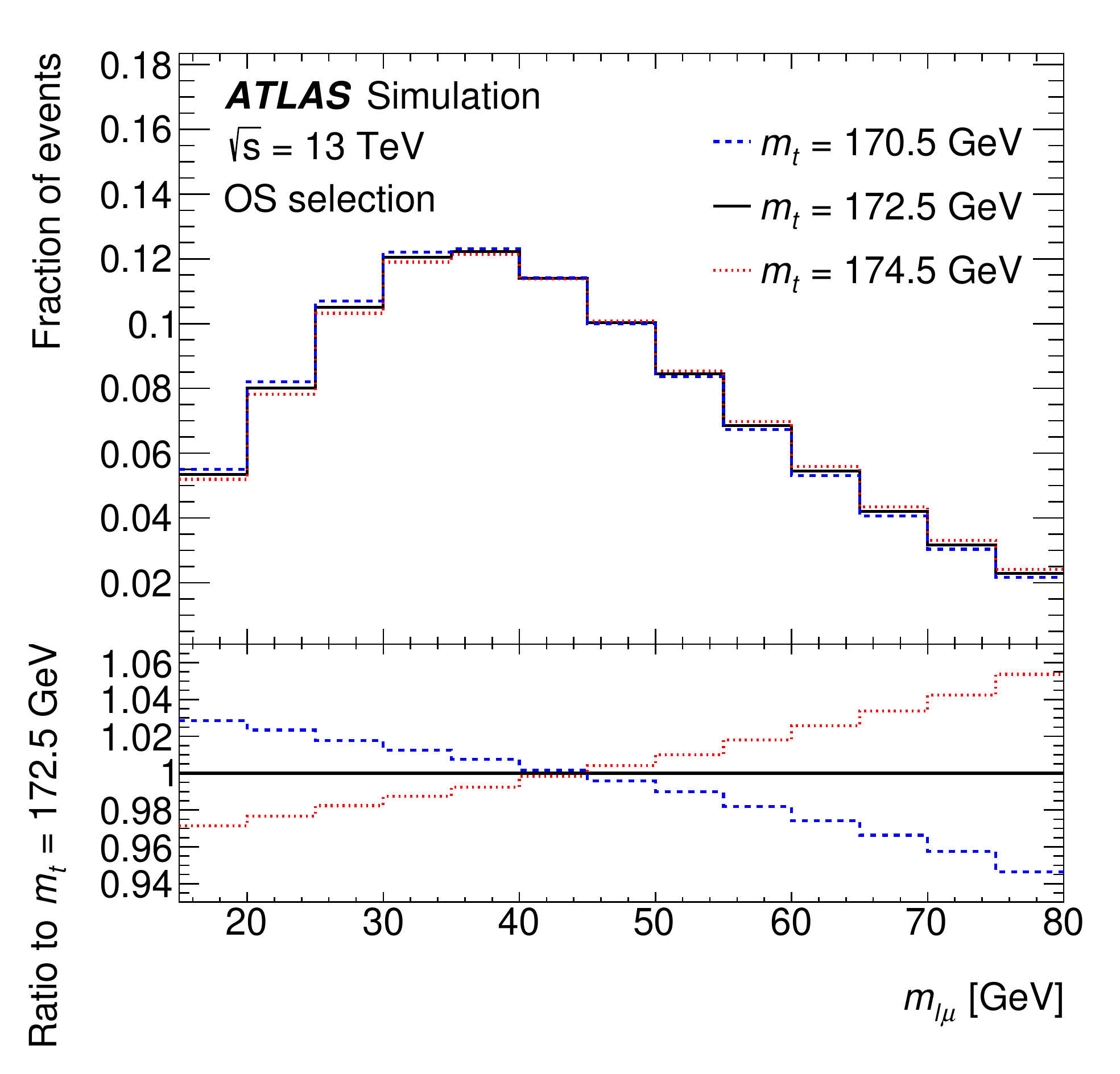}}
\subfloat[]{\includegraphics[width=0.45\textwidth]{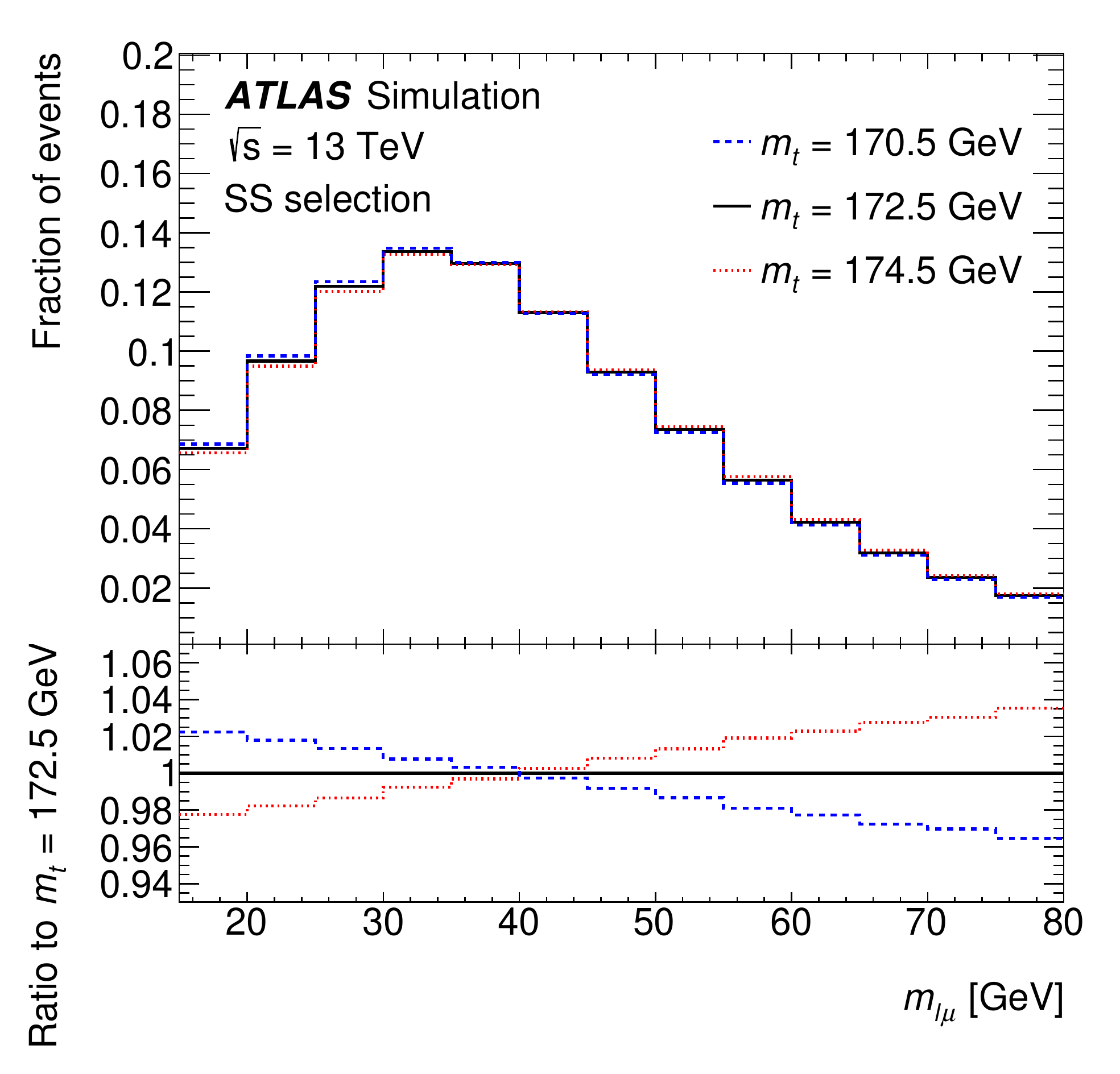}}
\caption{Sensitivity of the \mlmu{} distribution to different
input top-quark masses from simulated events, separately for the OS and SS samples.}
\label{fig:MassVariations_OS_SS}
\end{figure}

The distribution of the invariant mass of the primary lepton and
the soft muon, \mlmu{} is used to determine the mass of the
parent top-quark. A binned-template profile likelihood fit is
performed, with a Poisson likelihood model and systematic uncertainties included as Gaussian-constrained
nuisance parameters~\cite{HistFactory}. Only the range of \mlmu{} between 15 and 80~\GeV\
is considered in the fit, since the tail of the \mlmu{}
distribution is more sensitive to \ttbar modelling uncertainties and
higher-order corrections, and to the $Z$+jets background. The fit is
performed simultaneously for the OS and SS charge-combination samples,
and Figure~\ref{fig:MassVariations_OS_SS} shows the sensitivity of each of these
distributions to variations of the top-quark mass, as well as the binning used
by the templates. The SS sample has less sensitivity than
the OS sample due to the larger incidence of sequential
$b\rightarrow c\rightarrow \mu$ decays, where the SMT muon carries a
smaller fraction of the parent $b$-quark momentum, and due to the
larger fraction of events in which the leptons originate from
different top-quarks.
 
The fit uses template histograms simulated as for the nominal \ttbar
sample but with different values for the input top-quark mass.
The templates from the different mass samples are interpolated with
piece-wise linear functions built bin by bin. To improve the stability of the method, the templates are smoothed
assuming a linear dependence on $m_{t}$ for the fraction of the total number of
events in each bin. A maximum-likelihood fit is performed with three
free parameters: $m_t$, which controls the shape of the \mlmu{}
distribution for \ttbar\ events, and the normalisation factors for \ttbar events in the
OS and SS samples.
The normalisation factors ensure that the total of the \ttbar signal
and background events is always equal to the total number of selected data events,
and no $m_t$ information is extracted from the number of events.
 
The uncertainty due to the limited number of simulated events, and due
to statistical fluctuations in the background estimates based on control samples,
is evaluated by defining a new source of systematic uncertainty for
each bin of the prediction, which modifies the bin content by its statistical uncertainty.
Since a very large number of systematic uncertainties are considered
\textit{a priori}, a pruning procedure is applied to reduce the number of statistically
insignificant systematic uncertainties affecting the prediction of each of the signal and background
processes. A systematic variation of the \mlmu{} templates is excluded if the total predicted
change is smaller than 0.05\% of the nominal bin
content for all bins. The impact on the total estimated uncertainty is smaller than
0.03~\GeV.
 
\begin{figure}
\centering
\subfloat[]{\includegraphics[width=0.45\textwidth]{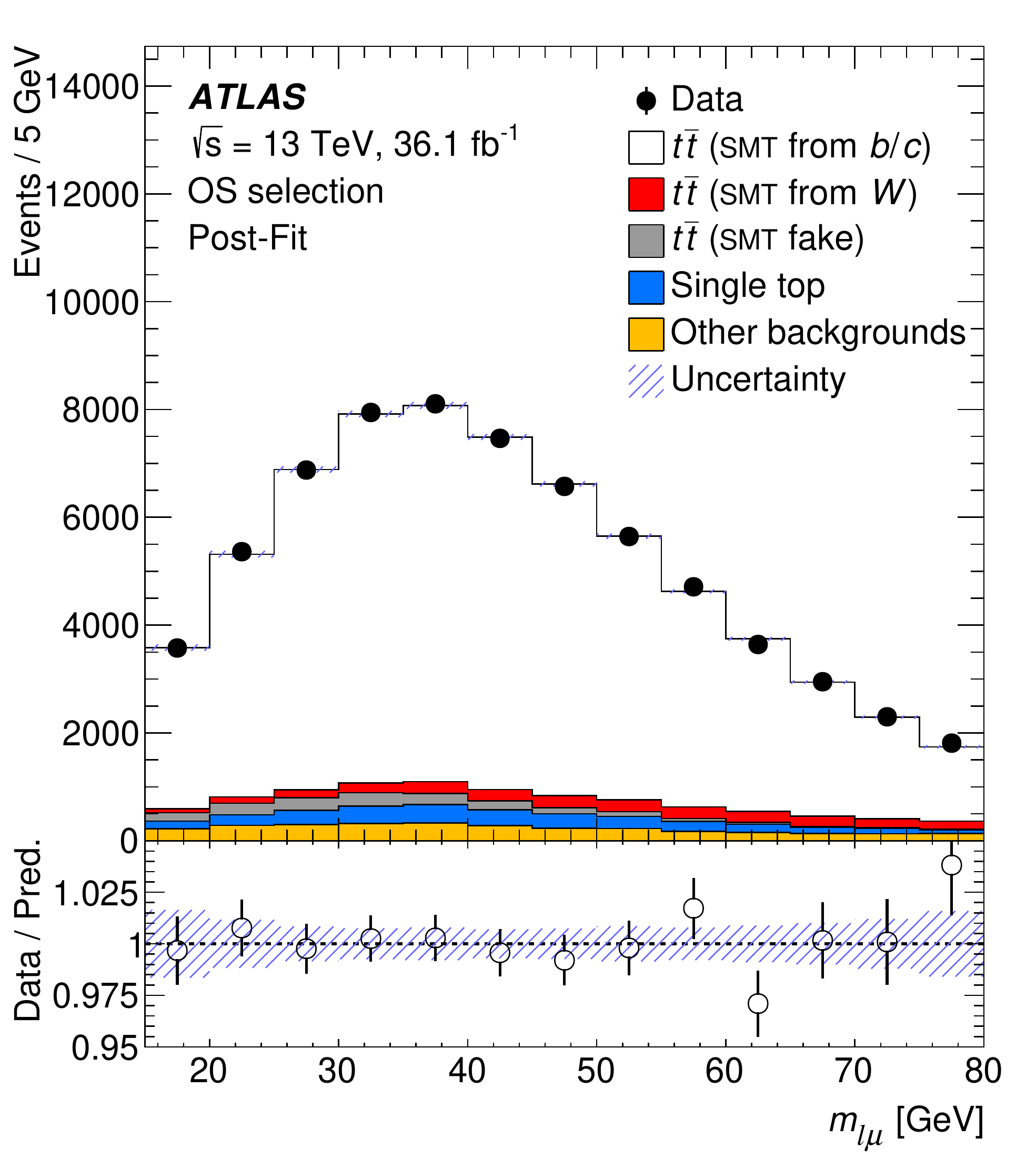}}
\subfloat[]{\includegraphics[width=0.45\textwidth]{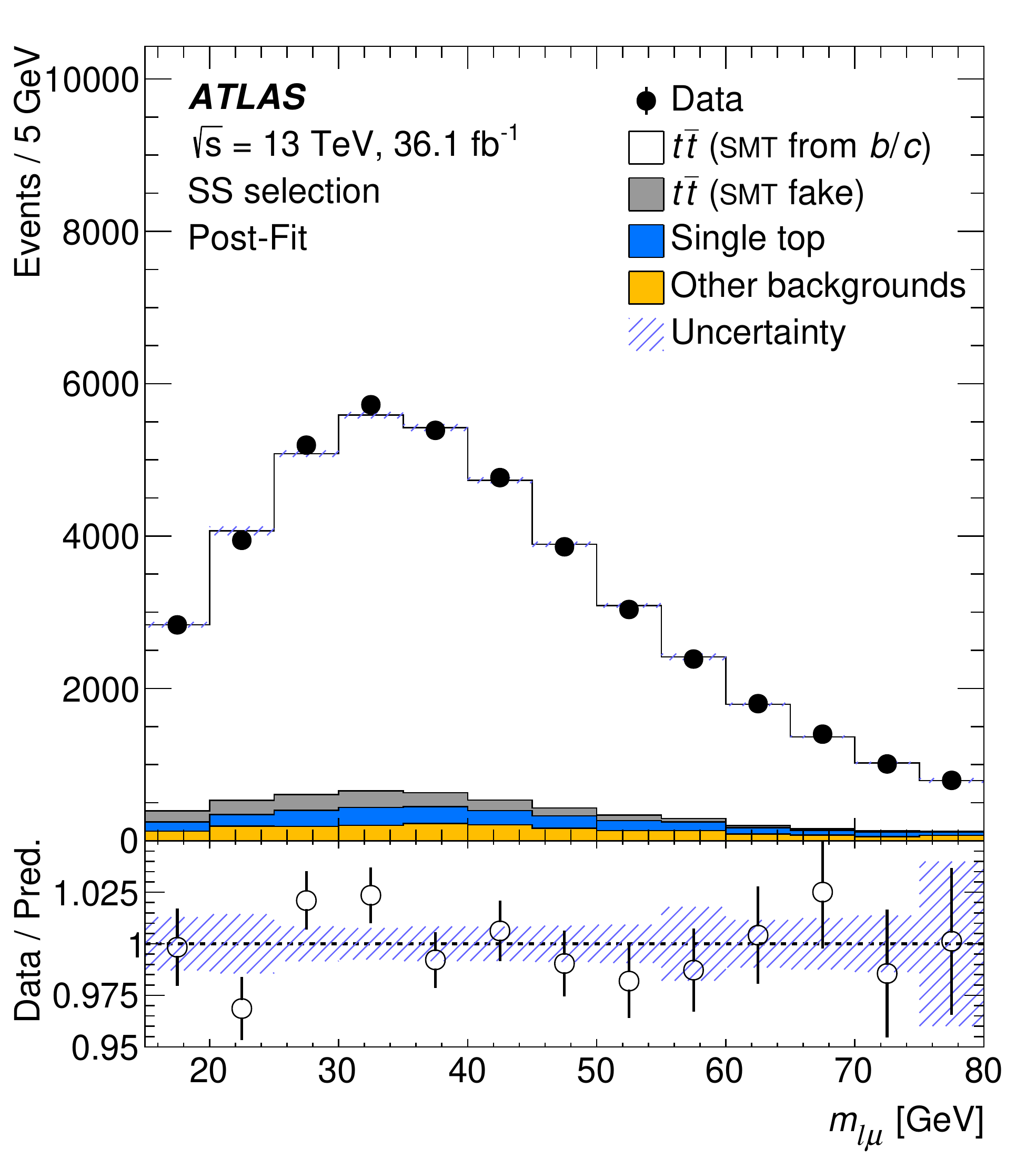}}
\caption{Post-fit \mlmu{} distributions in the (a) OS sample and
in the (b) SS sample. The prediction reports the event contribution for the
signal and backgrounds. The uncertainty band includes statistical
and systematic uncertainties, but does not include the recoil uncertainty.
The SS (same sign) or OS (opposite sign) refers to the charge signs of the primary lepton and the soft muon.}
\label{fig:finalfit}
\end{figure}

\begin{figure}[!h]
\centering
\subfloat[]{\includegraphics[width=0.45\textwidth]{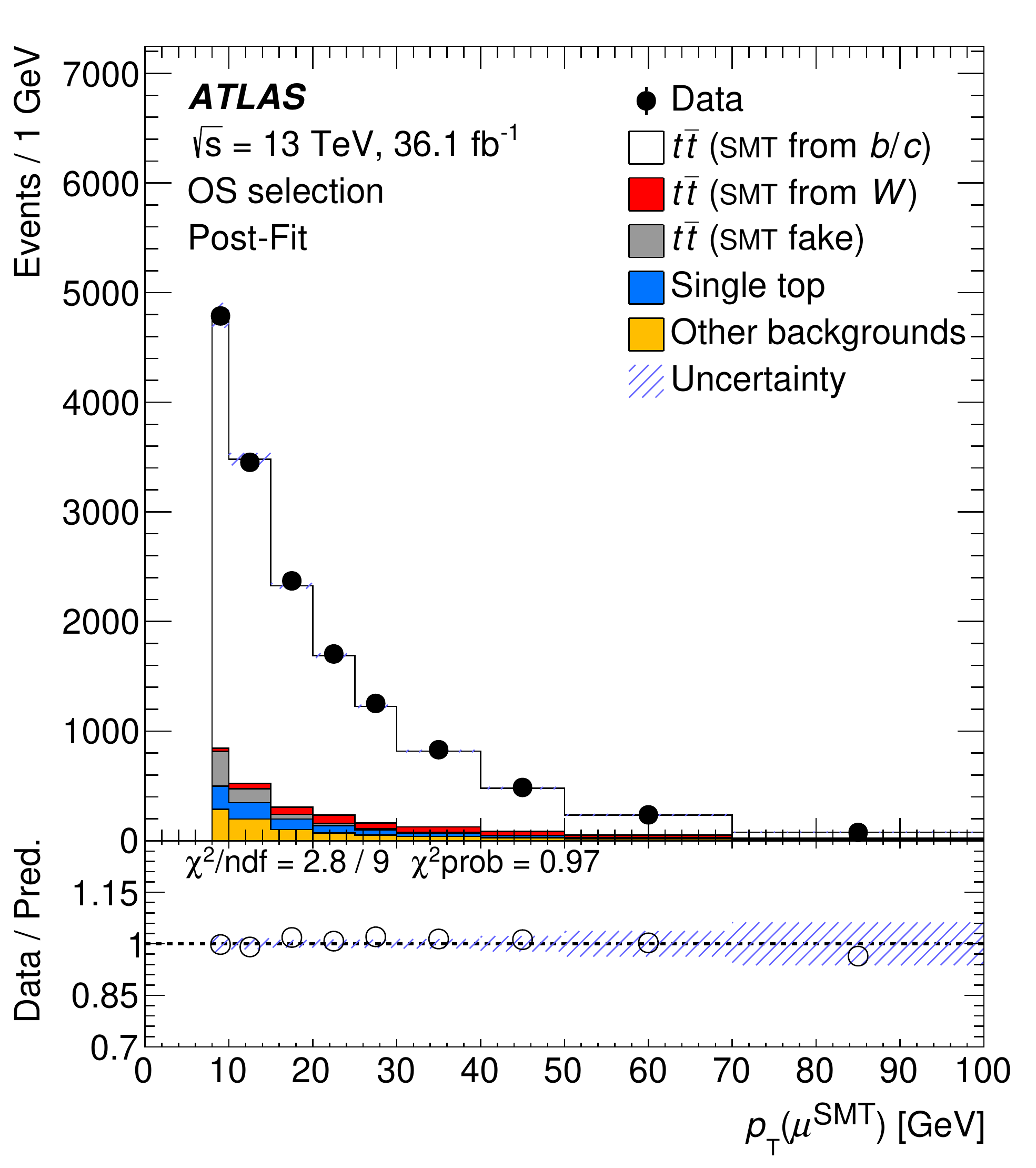}}
\subfloat[]{\includegraphics[width=0.45\textwidth]{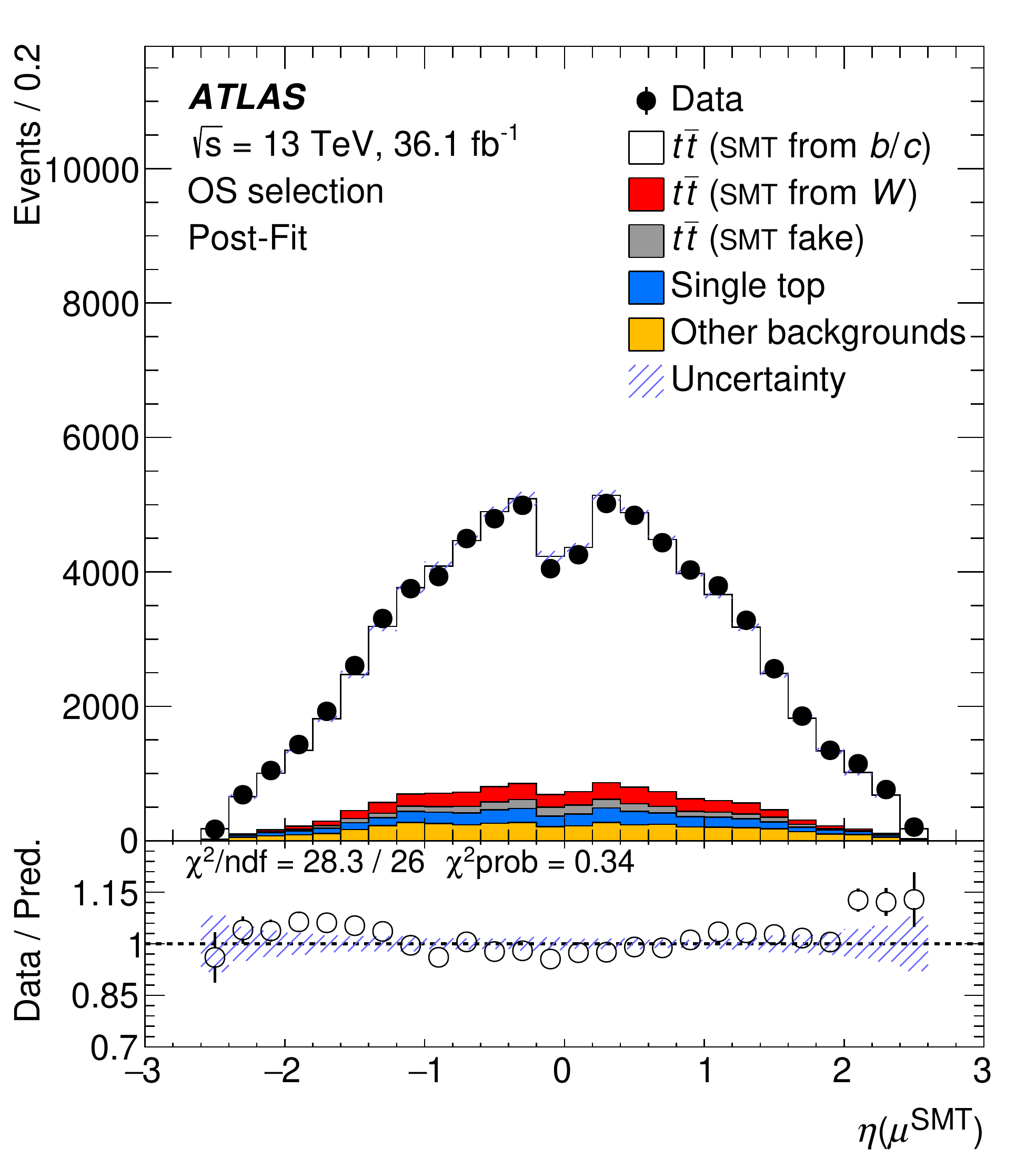}}
\\
\subfloat[]{\includegraphics[width=0.45\textwidth]{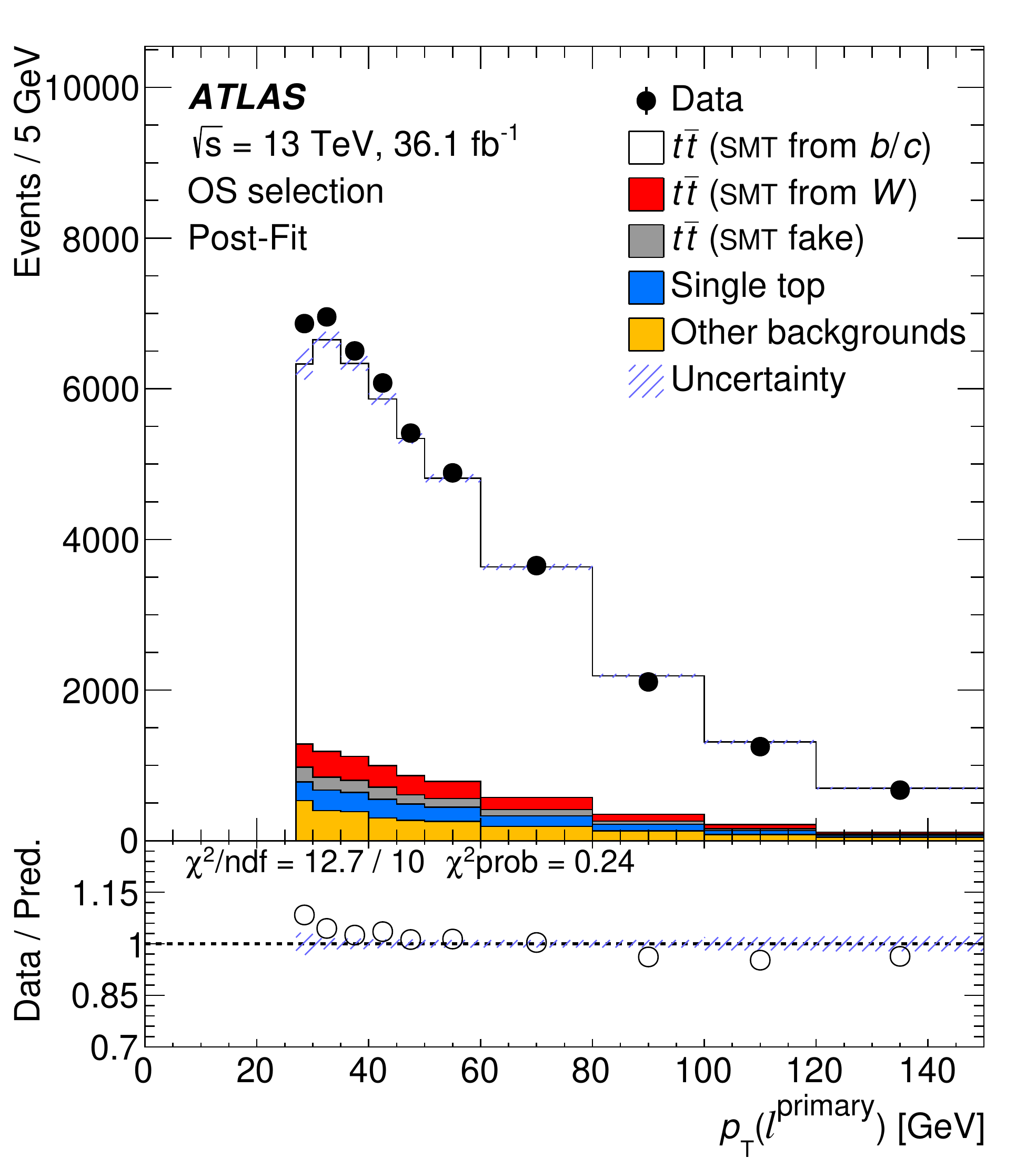}}
\subfloat[]{\includegraphics[width=0.45\textwidth]{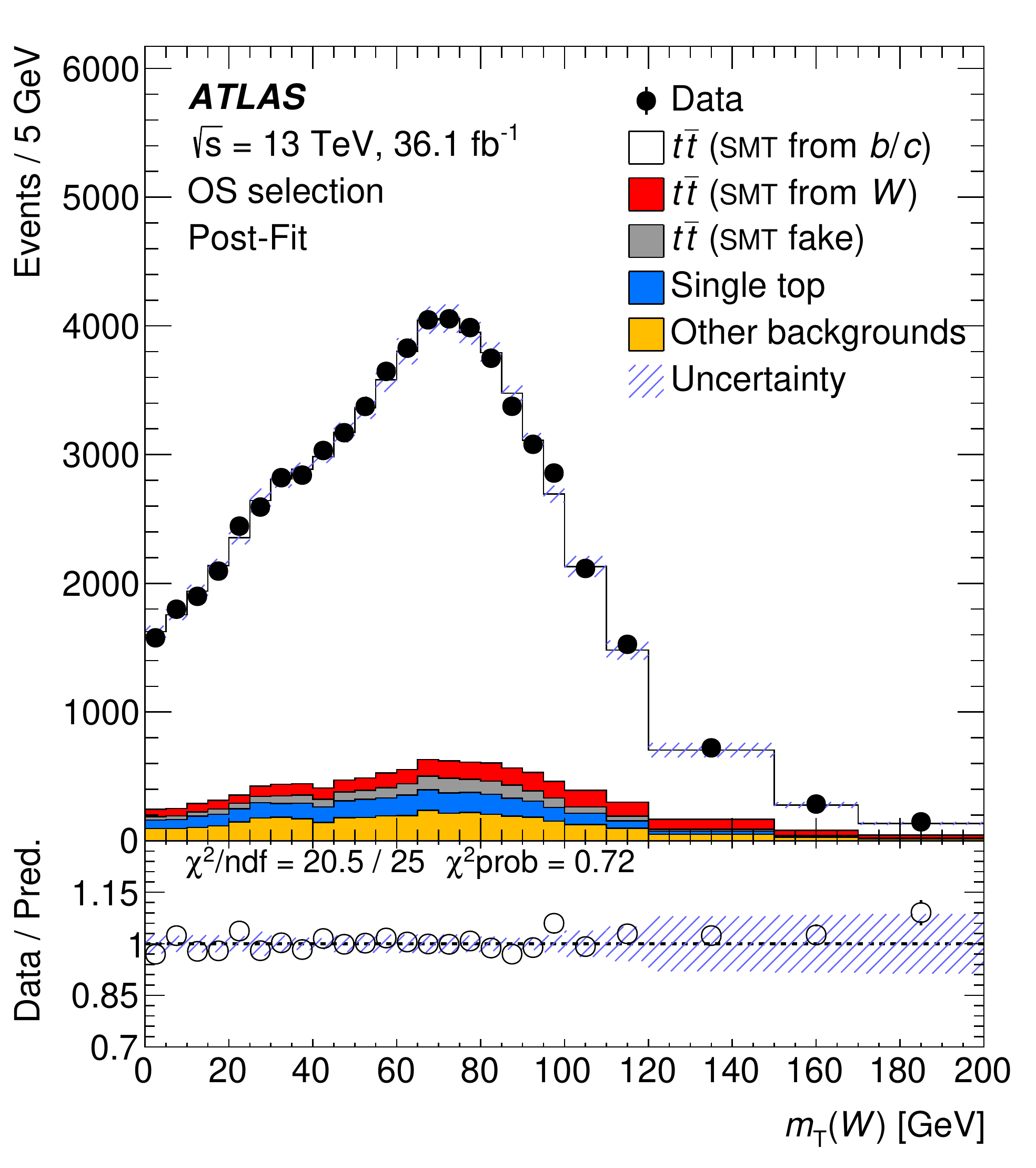}}
\caption{Post-fit comparison of data and prediction in the OS sample, i.e. for events with primary lepton and the soft muon with opposite charges, for the (a) soft muon \pt, (b) soft muon $\eta$,
(c) primary lepton \pt and (d) $W$-boson transverse mass. The prediction reports the event contribution
for the signal and backgrounds. The uncertainty band includes
statistical and systematic uncertainties, but does not include the recoil uncertainty.}
\label{fig:plots_bTagAnyOpti_Calib_postfit_OS}
\end{figure}

\begin{figure}[!h]
\centering
\subfloat[]{\includegraphics[width=0.45\textwidth]{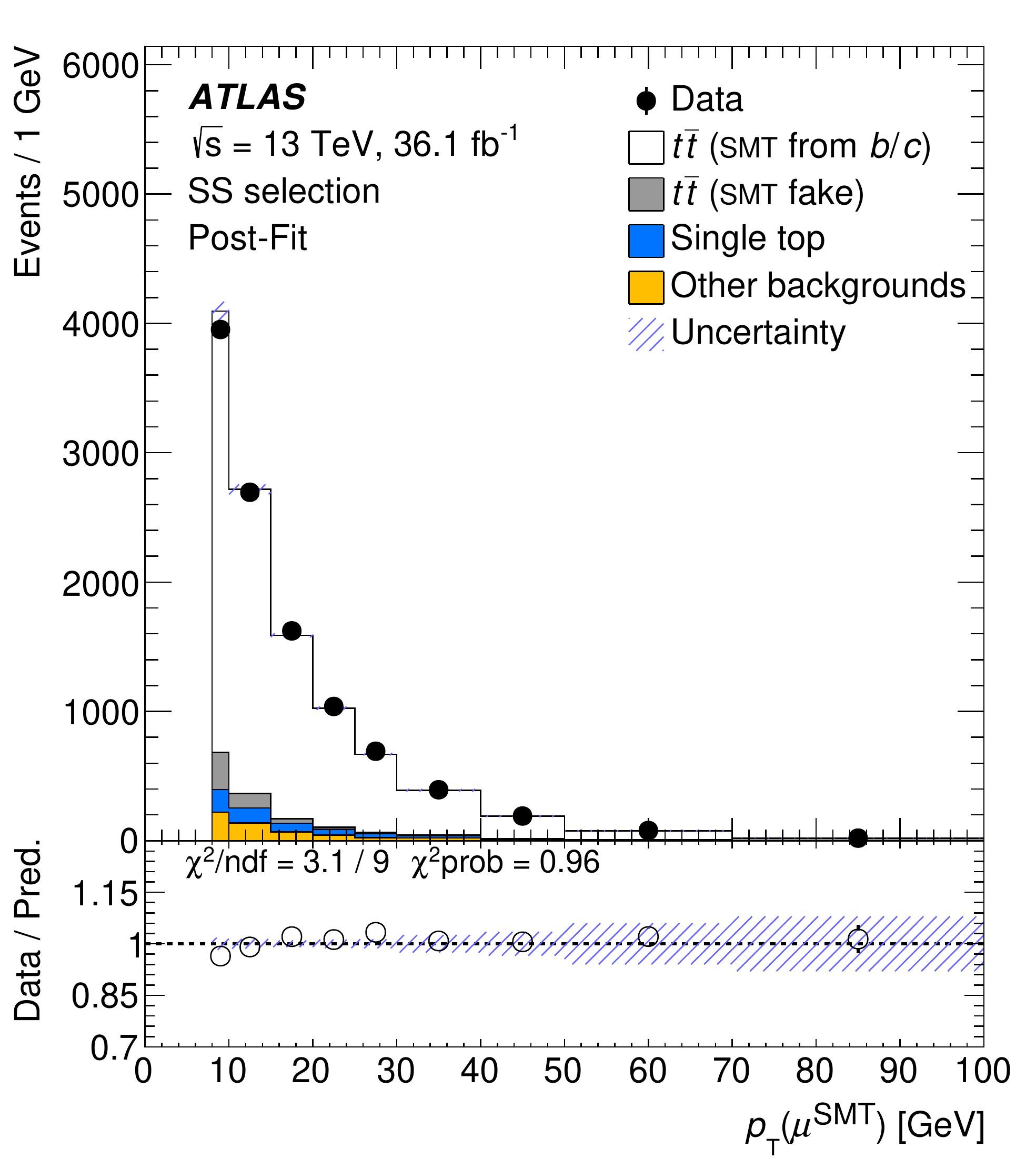}}
\subfloat[]{\includegraphics[width=0.45\textwidth]{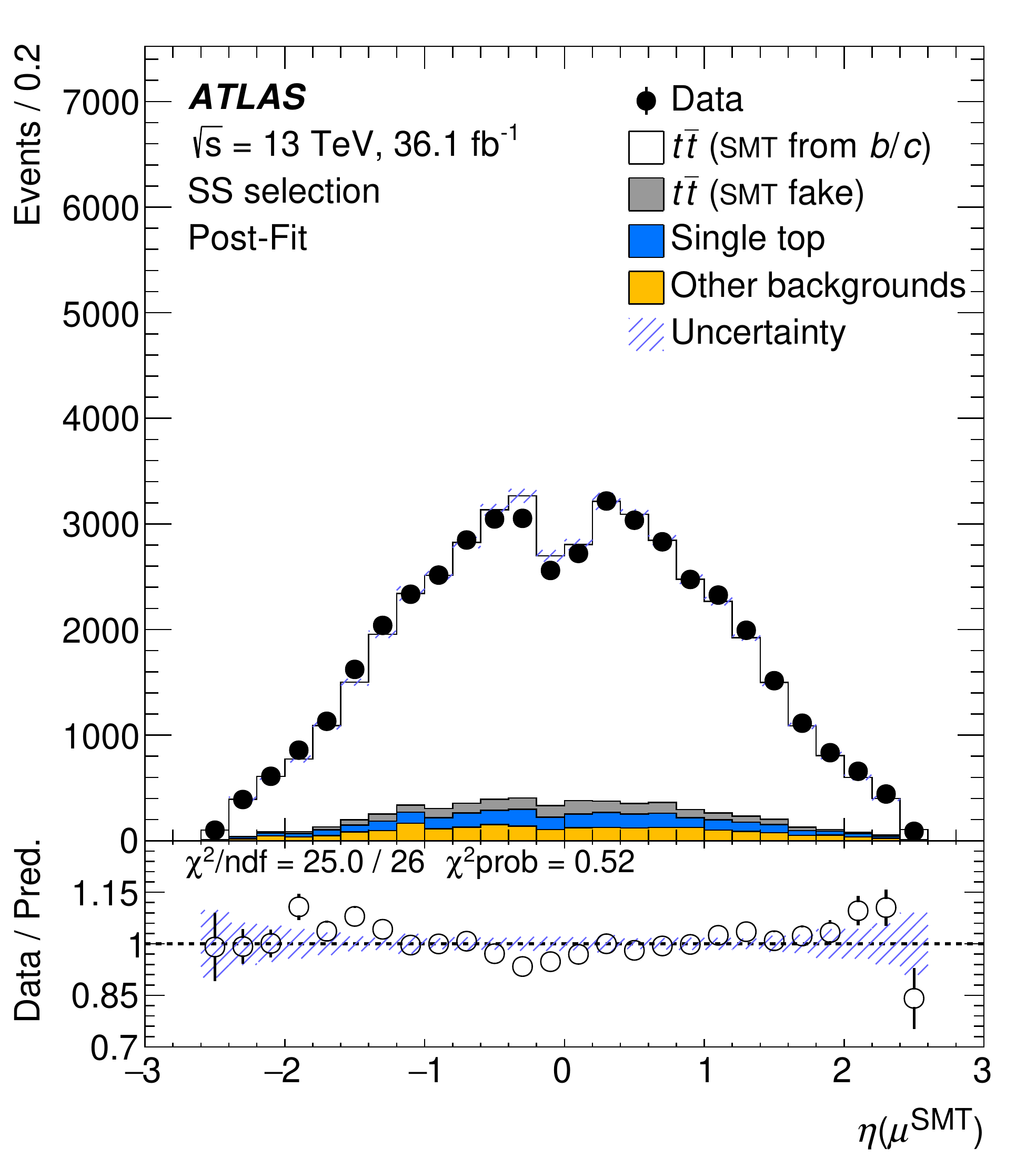}}
\\
\subfloat[]{\includegraphics[width=0.45\textwidth]{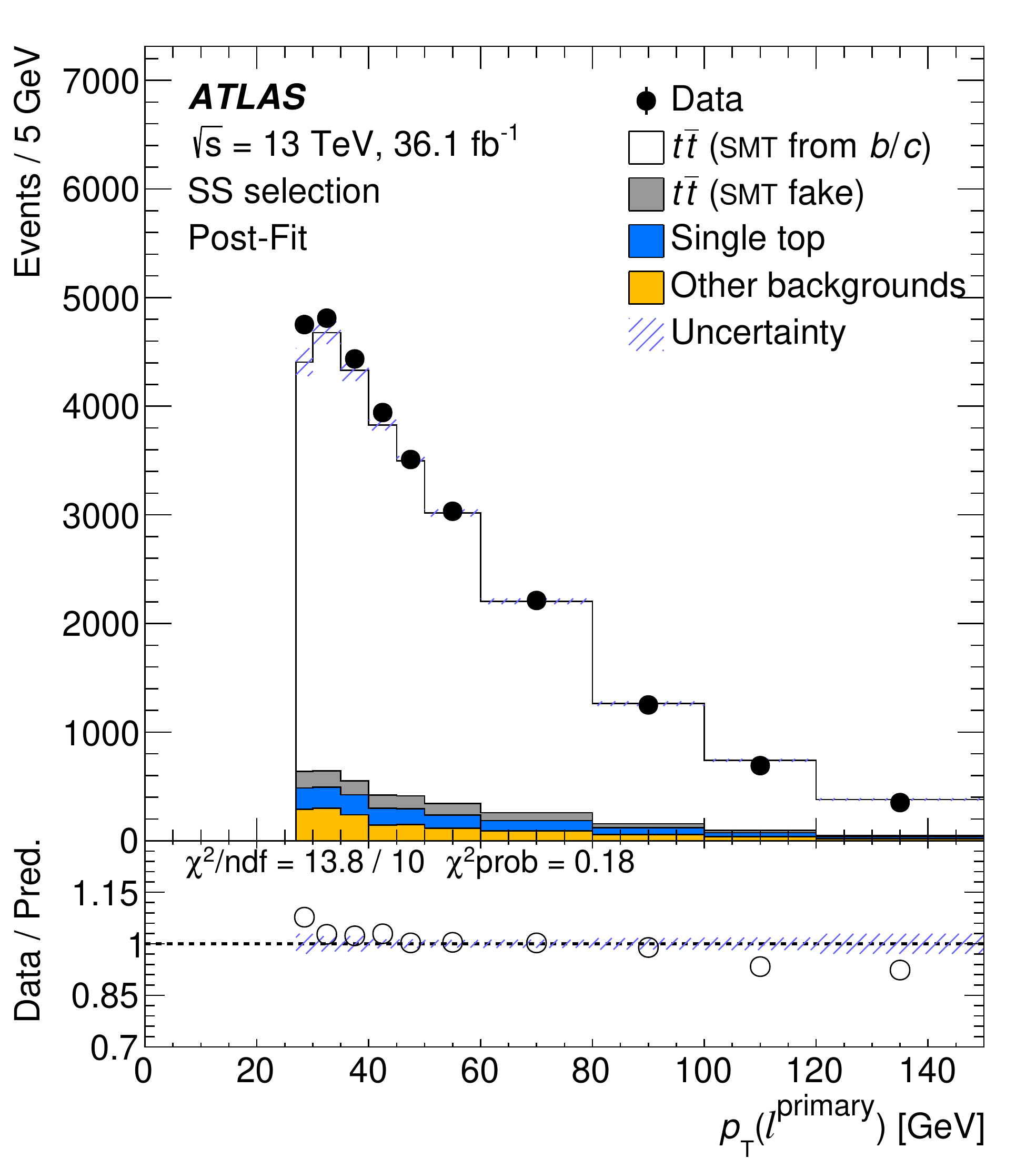}}
\subfloat[]{\includegraphics[width=0.45\textwidth]{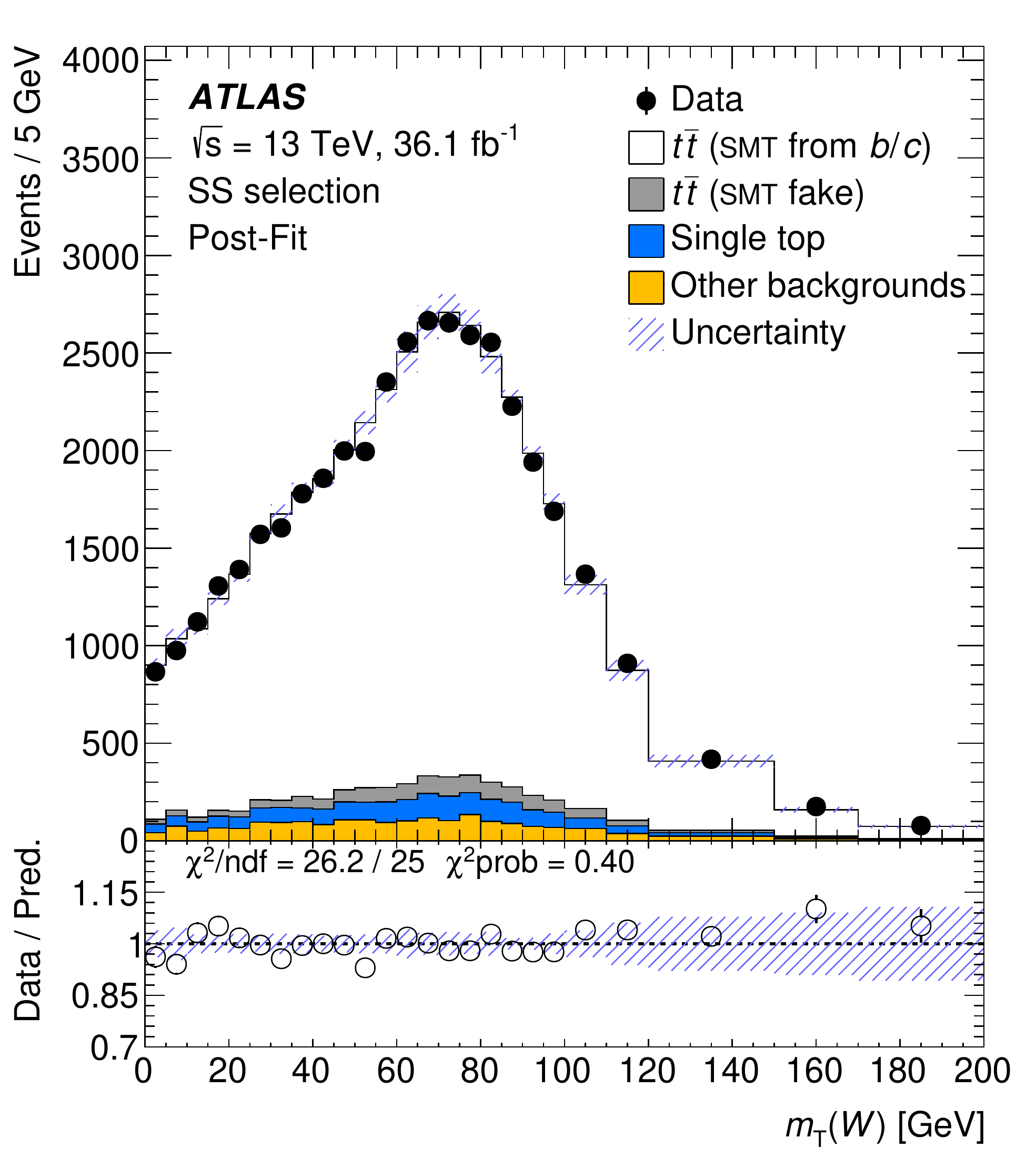}}
\caption{Post-fit comparison of data and prediction in the SS sample, i.e. for events with primary lepton and the soft muon with same charges, for the (a) soft muon \pt, (b) soft muon $\eta$,
(c) primary lepton \pt and (d) $W$-boson transverse mass. The prediction reports the event contribution
for the signal and backgrounds. The uncertainty band includes
statistical and systematic uncertainties, but does not include the recoil uncertainty.}
\label{fig:plots_bTagAnyOpti_Calib_postfit_SS}
\end{figure}

\begin{figure}
\centering
\includegraphics[width=0.58\textwidth]{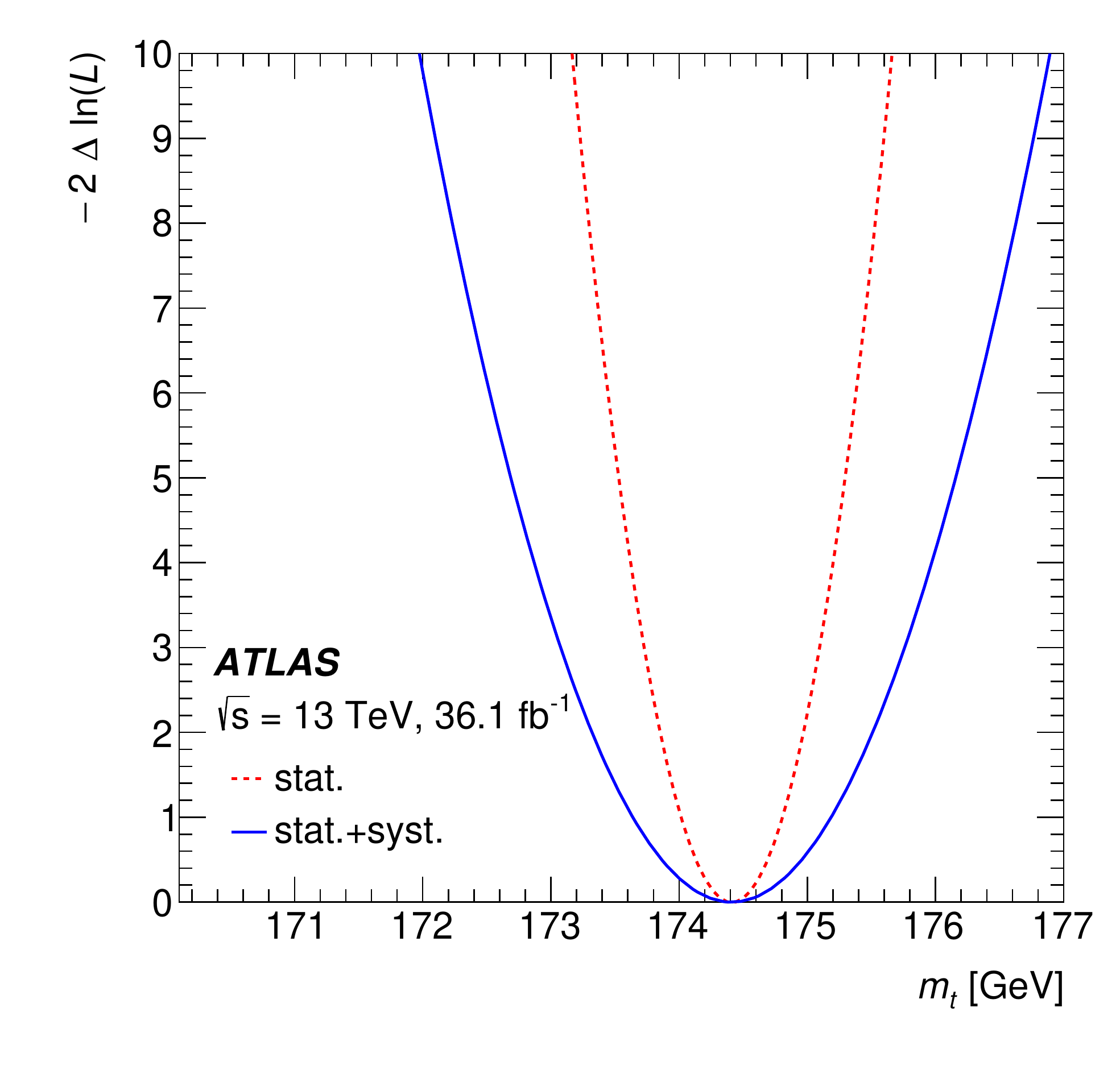}
\caption{Likelihood scan, showing the best-fit value and the
statistical and total uncertainty profiles (excluding the recoil uncertainty).}
\label{fig:fit_pulls_a}
\end{figure}

The top-quark mass determination from the fit is found to be linear
and unbiased with respect to the input top-quark mass hypothesis by means of pseudo-experiments, and its
uncertainty from the likelihood ratio is also checked to ensure it reports the
correct statistical coverage. The fit method and the event selection were optimised to minimise the
total uncertainty in $m_t$ in a `blinded' approach, i.e.\ using
pseudo-data and data without knowledge of the best-fit top-quark mass.
The fit yields:
 
\begin{equation*}
m_{t}
=174.41\pm0.39~(\textrm{stat.})\pm0.66~(\textrm{syst.})\pm0.25~(\textrm{recoil})~\textrm{GeV}, 
\end{equation*}
 
where the statistical, systematic and recoil uncertainties are described in detail in Section~\ref{sec:systematic_uncertainties}.
Figure~\ref{fig:finalfit} shows the post-fit \mlmu{}
distributions in the OS and SS samples; a goodness-of-fit test is performed using the saturated model
technique~\cite{Cousins1984,PDG:2018} and returns a probability of 56\%.
Figures~\ref{fig:plots_bTagAnyOpti_Calib_postfit_OS} and~\ref{fig:plots_bTagAnyOpti_Calib_postfit_SS} display the corresponding
post-fit plots for the kinematic variables of
Figures~\ref{fig:plots_bTagAnyOpti_Calib_OS} and~\ref{fig:plots_bTagAnyOpti_Calib_SS}.
The data distributions are well described by the prediction, with the
primary lepton \pt exhibiting a slight trend which is traced to
the boost of the \ttbar system, but which has no appreciable impact on
the determined top-quark mass.
This was confirmed by detailed checks, performed by testing the impact of NNLO corrections on the top quark kinematics, and by performing a test fit including the lepton pT as a second fit variable. In all cases the impact on the measurement
was shown to be well within the quoted modelling uncertainties associated with ISR effects and ME generator choice.
The post-fit uncertainties shown in Figures~\ref{fig:plots_bTagAnyOpti_Calib_postfit_OS} and~\ref{fig:plots_bTagAnyOpti_Calib_postfit_SS} are significantly reduced with respect to the pre-fit ones shown in Figures~\ref{fig:plots_bTagAnyOpti_Calib_OS} and ~\ref{fig:plots_bTagAnyOpti_Calib_SS} due to the \ttbar normalisation being
treated as a free parameter in the fit procedure; the normalisation uncertainty considered at pre-fit level is thus removed after the fit procedure.
The likelihood scan with the best-fit top-quark mass value is shown in Figure~\ref{fig:fit_pulls_a}.
 
Checks were performed by fitting the OS and SS regions separately,
giving  $m_{t}(\textrm{OS})=174.63\pm0.47~(\textrm{stat.})\pm0.75~(\textrm{syst.})~\GeV$ and
$m_{t}(\textrm{SS})=173.88\pm0.74~(\textrm{stat.})\pm1.01~(\textrm{syst.})~\GeV$.
Checks were performed by separately fitting the electron and muon channels, different $W$-decay lepton charges and different
configurations of $b$-tagging and event selection, and were found to
all give consistent results.
Checks also included the extraction of the top-quark mass with
alternative statistical methods, namely using analytic functions for
\mlmu{} with a parametric dependence on the top-quark mass, only using
the mean value of the \mlmu{}
distribution, and using a binned-template likelihood fit without
including systematic uncertainties as nuisance parameters. In
particular, the inclusion of systematic uncertainties as nuisance
parameters in the fit reduces the total uncertainty
by 2.6\%,  in line with reasonable constraints from the fit.
\section{Measurement uncertainties}
\label{sec:systematic_uncertainties}
The individual sources of uncertainty and the evaluation
of their effect on $m_t$ are described in the following.
Many sources of systematic uncertainty are considered, corresponding
to a total of 151 individual variations. Table~\ref{tab:fit_systs} summarises the impact on $m_t$ of the main
sources of systematic uncertainty, and each systematic uncertainty is accompanied by an estimate of its
statistical precision performed using the bootstrap method~\cite{ATL-PHYS-PUB-2021-011}.
\begin{table}
\centering
\caption{Impact of main sources of uncertainty on $m_t$.
Each row of the table corresponds to a group of individual systematic variations.
For each uncertainty source the fit is repeated with the corresponding group of nuisance parameters fixed to their best-fit values. The contribution from each source is then evaluated by subtracting in quadrature the uncertainty obtained in this fit
from that of the full fit. The total systematic uncertainty is different from the sum in quadrature of the different groups due to
correlations among nuisance parameters in the fit. The last column shows the statistical uncertainty on each of the top-quark mass uncertainties as estimated with the bootstrap method~\cite{ATL-PHYS-PUB-2021-011}. }
\begin{tabular}{lrr}
\toprule
Source & Unc. on $m_t$ [GeV] & Stat. precision [GeV] \\
\midrule

\textbf{Statistical and datasets} &   &  \\
\hspace{0.2cm} Data statistics & $0.39$ & \\
\hspace{0.2cm} Signal and background model statistics & $0.17$ & \\
\hspace{0.2cm} Luminosity & $<0.01$ & $\pm 0.01$ \\
\hspace{0.2cm} Pile-up & $0.07$ & $\pm 0.03$ \\
\midrule
\textbf{Modelling of signal processes} &   &  \\
\hspace{0.2cm} Monte Carlo event generator & $0.04$ & $\pm 0.06$ \\
\hspace{0.2cm} $b,c$-hadron production fractions & $0.11$ & $\pm 0.01$ \\
\hspace{0.2cm} $b,c$-hadron decay BRs & $0.40$ & $\pm 0.01$ \\
\hspace{0.2cm} $b$-quark fragmentation $r_b$ & $0.19$ & $\pm 0.06$ \\
\hspace{0.2cm} Parton shower $\alpha_S^{FSR}$ & $0.07$ & $\pm 0.04$ \\
\hspace{0.2cm} Parton shower and hadronisation model & $0.06$ & $\pm 0.07$ \\
\hspace{0.2cm} Initial-state QCD radiation & $0.23$ & $\pm 0.08$ \\
\hspace{0.2cm} Colour reconnection & $<0.01$ & $\pm 0.02$ \\
\hspace{0.2cm} Choice of PDFs & $0.07$ & $\pm 0.01$ \\
\midrule
\textbf{Modelling of background processes} &   &  \\
\hspace{0.2cm} Soft muon fake & $0.16$ & $\pm 0.03$ \\
\hspace{0.2cm} Multijet & $0.07$ & $\pm 0.02$ \\
\hspace{0.2cm} Single top & $0.01$ & $\pm 0.01$ \\
\hspace{0.2cm} $W$/$Z$+jets & $0.17$ & $\pm 0.01$ \\
\midrule
\textbf{Detector response} &   &  \\
\hspace{0.2cm} Leptons & $0.12$ & $\pm 0.01$ \\
\hspace{0.2cm} Jet energy scale & $0.13$ & $\pm 0.02$ \\
\hspace{0.2cm} Soft muon jet p$_\textrm{T}$ calibration & $<0.01$ & $\pm 0.01$ \\
\hspace{0.2cm} Jet energy resolution & $0.08$ & $\pm 0.07$ \\
\hspace{0.2cm} $b$-tagging & $0.10$ & $\pm 0.01$ \\
\hspace{0.2cm} Missing transverse momentum & $0.15$ & $\pm 0.01$ \\
\midrule
\midrule
Total stat. and syst. uncertainties (excluding recoil) & $0.77$ & $\pm 0.03$ \\
\midrule
Recoil uncertainty & $0.25$ & \\
\midrule
Total uncertainty & $0.81$ & \\
\bottomrule


\end{tabular}
\label{tab:fit_systs}
\end{table}
 
\subsection{Statistical and datasets} The uncertainty related to the size of the data sample (data statistical uncertainty) is
obtained by performing the fit while keeping constant all of the nuisance
parameters associated with the systematic uncertainties. The data
statistical uncertainty obtained by using both the OS and SS
selections is $\pm$0.39~GeV, while the OS sample alone yields
$\pm$0.47~GeV, highlighting the contribution to the sensitivity from
the SS sample.
The uncertainty due to the limited size of the simulated signal
and background samples
includes both the impact of the signal samples size on $m_t$-dependent templates used for the top mass interpolation, and the uncertainty of the backgrounds due to the limited size of the corresponding MC samples.
This includes the multijet background, which is estimated with
a data control sample.  The uncertainty in the combined 2015--2016 integrated luminosity is
2.1\%~\cite{ATLAS-CONF-2019-021}, obtained using the LUCID-2
detector~\cite{LUCID2} for the primary luminosity measurements.
The distribution of the average number of interactions per bunch
crossing (pile-up activity) in each MC sample is reweighted to match
the conditions in data, and a corresponding uncertainty is evaluated according
to the uncertainty in the average number of interactions per bunch crossing.
 
\subsection{Modelling of the signal process}
Uncertainties in the  \ttbar signal modelling include all sources that affect the
kinematics of the lepton from the $W$-boson decay
and the kinematics of the $b$-hadron giving rise to the soft muon, and
also the fraction of events from different soft-muon flavour components (from $b$-hadrons, $c$-hadrons,
light jets and $W$-bosons).
The \ttbar\ inclusive cross-section uncertainty does not affect the measurement,
since the fit is based only on the shape of the distribution from \ttbar{} events after background subtraction.
 
Uncertainties that depend on the choice of NLO matching scheme in the \ttbar{} MC generator are estimated
by comparing a sample generated with \textsc{Powheg+Pythia8}
with a sample generated with
\textsc{MadGraph5}\_a\textsc{MC@NLO+Pythia8} \cite{Alwall:2014hca} (referred to hereafter as a\textsc{MC@NLO+Pythia8}).
The a\textsc{MC@NLO} matching requires specific settings of the \textsc{Pythia8} shower
to retain the NLO accuracy. The matrix-element corrections are switched off for both initial-state radiation
and the global-recoil settings that are used for final-state radiation
emissions. These settings are different from the nominal
\textsc{Powheg+Pythia8}. In order to have a coherent comparison,
an alternative \textsc{Powheg+Pythia8} sample was generated
with the same \textsc{Pythia8} configuration as that used to shower a\textsc{MC@NLO} events.
Additionally, since a\textsc{MC@NLO+Pythia8} is known to describe
poorly the distribution of the boost of the \ttbar
system ($p_\textrm{T}^{t\bar{t}}$)~\cite{TOPQ-2018-15}, the
$p_\textrm{T}^{t\bar{t}}$ in a\textsc{MC@NLO+Pythia8} is reweighted to that of the \textsc{Powheg+Pythia8} sample.
The full difference between the top-quark masses obtained with the two
samples is considered as the positive and negative uncertainty due to
the MC generator NLO matching.
 
Uncertainties in the $b$-hadron production fractions and the BRs of
the inclusive decays of $b$- and $c$-hadrons into muons are derived from the uncertainties in the
rescaling procedure, described in Section~\ref{sec:hq-fragmentation} and
shown in Tables~\ref{tab:prodfracPyt} and \ref{tab:BRs}. These uncertainties are propagated through the analysis.
In addition, a check was performed to verify that the impact on \mlmu\ due to the different admixture of $D$-mesons involved in $b \to c \mu + X$ transitions was within the uncertainty assigned to $b \to \mu$ inclusive BRs. For this purpose, the exclusive decays $B^0\rightarrow D^-\mu\nu$, $B^0\rightarrow D^*(2010)^-\mu\nu$, $B^+\rightarrow \overline{D^0}\mu\nu$, $B^+\rightarrow \overline{D^*}(2007)^0\mu\nu$ and their charge conjugates were considered. For each of these decays, a kinematic selection similar to that of the main selection was applied to the events and the BR of each decay was varied within the uncertainty quoted by the PDG. The impact on \mlmu{} was found to be significantly smaller than the effect of varying only the BR of the inclusive $b$ $\to$ $\mu$ decays.
The impact on \mlmu{} from the uncertainties in the $B^0_{(s)}$ mixing parameters is
much smaller than the impact from imperfect knowledge of the $b$-hadron production
fractions and from the heavy-quark hadrons BRs.
Similarly, the impact of the modelling of the soft-muon kinematics in the exclusive semileptonic decays of $b$- and $c$-hadrons in \textsc{EvtGen}~v1.2.0 was tested by varying the various BRs within their uncertainties. The total impact was found to be negligible with respect to the impact of the inclusive $b$-hadron production fractions and heavy-quark hadrons BRs uncertainties.
 
Uncertainties in the modelling of the parton shower and
hadronisation processes include the estimation of several components.
An alternative simulation of the \ttbar\ sample is
considered whereby the \textsc{Powheg-Box} generator is matched
to the \textsc{Herwig}~7.1.3 generator for the modelling of the parton shower and
hadronisation.  The \textsc{Herwig}~7.1.3 generator release includes several
improvements in the shower description for heavy-quark fragmentation, together with a new tune to $e^+e^-$ data.
The angle-ordered shower algorithm is preferred to the dipole shower for this sample, as it better
describes both the shower evolution in the 7~\TeV{} ATLAS measurement of jet shapes
in \ttbar{} events~\cite{STDM-2011-48}, and the $x_B$ distribution
of LEP data, although it does not describe the $x_B$ spectrum of LEP data as
well as \textsc{Pythia8}.
The sample based on \textsc{Herwig}~7.1.3, when compared with the nominal \ttbar simulation used in the
fit, allows the effects of changes in the shower algorithm, and therefore in initial- and final-state emissions, to
be assessed using alternative but coherent models of $b$-quark fragmentation and hadronisation, the
underlying event and colour reconnection.
The full difference between the top-quark masses obtained with the
two samples  of \textsc{Powheg} events showered with \textsc{Pythia8}
A14-$r_b$ and  \textsc{Herwig}~7.1.3 is considered as the positive and
negative uncertainty from variations of the parton shower and hadronisation
modelling.
 
To evaluate the uncertainty on the modelling of the $b$-quark fragmentation, additional samples were produced
with the value of $r_b$ in the fragmentation function varied by its
uncertainty of $\pm 0.02$.
Additionally, the uncertainty on the choice of the renormalisation and factorisation scales in the final-state radiation (FSR) description of \textsc{Pythia8} is evaluated using alternative simulated \ttbar samples generated with the scale parameters explicitly\footnote{
With explicit scale variations, dedicated alternative MC samples are generated with the $\mu_{R,F}$ scales modified in the parton shower settings, as opposed to the automated
parton shower variations discussed in Ref.~\cite{PhysRevD.94.074005}, where weights are applied to the MC sample to obtain the systematic shifts. Explicitly changing $\mu_{R,F}$  by a factor $\sqrt{2}$ corresponds approximately to an automated variation of a factor 2, thanks to the implementation of the NLO compensation terms in the latter case.
}
varied up and down by factors of $\sqrt{2}$~\cite{MrennaSkands2016}. For these alternative settings, the appropriate $r_b$ values are determined following the same fit procedure described in Sec.~\ref{sec:hq-fragmentation}, in order to still correctly model the $x_B$ distribution for LEP/SLD data. It must be noted that the $x_B$ distribution will however be different for the \ttbar samples with varied scales with respect to the nominal one, due to process-dependent effects. This uncertainty is labelled $\alpha_S^{FSR}$ in Tab.~\ref{tab:fit_systs}.
 
As a check of the separate $r_b$ and $\alpha_\textrm{S}^\textrm{FSR}$
uncertainties reported in Table~\ref{tab:fit_systs}, a sample of
\textsc{Powheg} events showered with \textsc{Pythia} using the \textsc{Monash}
tune~\cite{Skands:2014pea} (which has different values of $\alpha_\textrm{S}$
and $r_b$ than the A14-$r_b$ tune), is used to obtain an equivalent systematic
uncertainty. The change in the measured top-quark mass obtained with
this sample is $0.30\pm0.06$~\GeV, consistent with the uncertainties associated with $\alpha_\textrm{S}^\textrm{FSR}$, $r_b$ and the hadronisation model, listed in Table~\ref{tab:fit_systs}.
 
In the modelling of the parton shower of the $b$-quark
from $t\rightarrow Wb$ with \textsc{Pythia} 8.2,
there is the possibility to change the default gluon recoil scheme
from recoiling against the $b$-quark (the nominal setting, referred to here as RTB), to recoiling
against the $W$-boson (\textsc{recoilToColoured=off}, referred to as
RTW)~\cite{SkandsRecoil}. Before \textsc{Pythia} version 8.160, the RTW was the only possibility,
but it could give unphysical radiation patterns and it is now kept as
an option to understand the effect this setting has in view of previous measurements.
This setting changes the modelling of second and subsequent gluon
emission from quarks produced by coloured resonance decays, such as
the $b$-quark in a $t\rightarrow Wb$ process, but it has no impact for
example on $Z\rightarrow b\bar{b}$ decays. A third recoil scheme has been recently made available via the
\textsc{UserHook} functionality of \textsc{Pythia} 8.2 with the
top-quark itself serving as recoiler for second and subsequent gluon emission of the
$b$-quark (referred to as RTT)\footnote{The code used for the RTT \textsc{UserHook} was provided by the \textsc{Pythia} authors directly.}. The RTW and RTT setups give
wider-angle gluon radiation, resulting in energy deposits that do not
get clustered into the $b$-jet, and lower gluon-energy emission, altering the modelling of the
$b$-quark fragmentation and hardening the $b$-hadron momenta.
They also mildly change the $W$-boson p$_T$ and the angle between the
$W$-boson and the $b$-hadron resulting from the top-quark decay.
The recently-developed RTT option has been considered as an additional
uncertainty in the measurement, even though the implementation could only be performed based on particle-level simulation and without a dedicated tune that would normally accompany a change of setup of this nature.
The change of the recoil model modifies the distribution of the momentum fraction of the $b$-hadron
\begin{equation*}
\label{eq:XB}
x_B=\frac{1}{1-m_W^2/m_t^2+m_b^2/m_t^2}\frac{2 {p}_B\cdot{p}_t}{m_t^2},
\end{equation*}
 
where $m_w$ is the $W$-boson mass and $p_t$ is the top four-momentum. However, the Mellin moments of this distribution derived with the RTB setup
agree well with those predicted by the NLO+NLL
resummation convoluted with the Kartvelishvili model tuned on ALEPH, OPAL and SLD
data~\cite{Corcella:2005}. Therefore, the $x_B$
distribution derived with the RTT shower is reweighted to that of the
RTB simulation. The reweighting of $x_B$ is a crude adjustment, since
its distribution is controlled by the parameters of the fragmentation
function and $\alpha_S^{FSR}$, which are set in the A14-$r_b$ \textsc{Pythia}
tune, and bound by the universality of the $b$-quark fragmentation
model. Closure tests demonstrated that the $x_B$ reweighting provides
the correct $m_{\ell \mu}$ distribution, and the correct extracted
top mass, even though it does not fully address the modelling of other variables of the event.
Overall, the analysis performed with the RTT shower setup as above
yields a measured top-quark mass value which is larger than the one
obtained with the nominal parton shower configuration by
approximately 0.25 GeV. The less physically-motivated RTW setup has
been also checked, and would have an effect at the same level as RTT.
A one-sided shift of the measured top mass is seen with the RTT and RTW setups, but
a symmetrised uncertainty of $\pm$0.25~GeV is assigned and indicated
by ``recoil''. This uncertainty is considered outside of the profile likelihood fit, and quoted as a separate uncertainty on the result.

The uncertainty in the modelling of initial-state radiation (ISR) is estimated by using four variations: the first two are obtained by independently changing the \textsc{Powheg} ISR renormalisation and factorisation scales
up and down by a factor 2.0, the third by comparing with an alternative sample obtained doubling the $h_\textrm{damp}$ parameter with respect to the nominal settings, the fourth one corresponds to the \texttt{Var3c} up and down variations of the \textsc{Pythia8} A14 tune~\cite{ATL-PHYS-PUB-2016-004}.

The modelling of the underlying event and of colour reconnection (CR) can affect the amount of radiation emitted from the $b$-quark,
as well as modify the kinematic distribution of the $b$-hadron. An
underlying-event uncertainty is estimated
using the corresponding eigentunes of the A14 \textsc{Pythia8} tune.
Variations of colour reconnection parameters are also provided by the A14 eigentunes,
determined from measurements of the underlying event in jet production.
Samples are generated where the colour reconnection strength in the \textsc{Pythia8} default model
is set to its maximum value (all hadrons are reconnected) and are compared
with a setting with the colour reconnection switched off. To account for
the possibility of colour reconnection also affecting the top-quark decay products,
a comparison with the `Early Resonance Decay' (ERD) model is performed~\cite{Argyropoulos:2014zoa}.
In this model, the top-quarks and $W$-bosons are allowed to decay before CR
takes place, so the top-quark decay products directly
participate in CR. The impact on the measured top-quark mass is found
to be negligible.
 
The systematic uncertainty due to the choice of PDFs is evaluated
using the \textsc{PDF4LHC15} error set~\cite{PDF4LHC15} applied to the nominal \ttbar{}
MC settings, and is obtained by means of event reweighting for 30
PDF replicas. The $m_t$ value is extracted for each of the 30 cases, and the total
systematic uncertainty due to this effect is computed as the sum in quadrature of the single variations.
 
\subsection{Modelling of background processes}
Several sources of uncertainty are considered for the normalisation
and shape of the background contributions.
The relative uncertainty in the normalisation of the soft-muon
component of \ttbar events which arises from light-hadron decays and
detector background (`Soft muon fake')
is derived from the uncertainty in the calibration of the misidentification rate.
An additional uncertainty is derived from the difference in shape and normalisation of the
\mlmu{} distribution of SMT mistags between \ttbar samples generated with
\textsc{Powheg+Pythia8} and \textsc{Sherpa}. For the \ttbar dilepton
component, uncertainties in the modelling of the ISR,
in the NLO matching and in the parton shower and hadronisation model are
estimated in the same way as for the nominal \ttbar\ sample.
 
An uncertainty of $+5\%/-4\%$ is applied to the total cross-section
for  single top-quark production~\cite{Kidonakis:2011wy,Kidonakis:2010ux,Kidonakis:2010tc}.
An additional uncertainty in the amount of initial- and final-state radiation is
evaluated in a manner similar to
that used for \ttbar. The uncertainty in the interference between $Wt$ and \ttbar\
production at NLO is assessed by comparing the default `diagram removal' scheme with an alternative
`diagram subtraction' scheme~\cite{Frixione:2008yi}. The uncertainty in
the event generator for the $t$-channel is evaluated by comparing it with
a sample simulated with a\textsc{MC@NLO+Pythia8}. The uncertainty in
the parton shower and hadronisation models for the $t$- and $Wt$-channels
is derived from a comparison with samples showered with
\textsc{Herwig++}. The impact of $m_{t}$ variations on the single top-quark background has been found to yield negligible effects.
 
An uncertainty of 30\% is applied to the $Z$+jets background
normalisation, for both its light-flavour-jet and
heavy-flavour-jet ($Z$+$c\bar{c}$ and $Z$+$b\bar{b}$) components.
It was validated in a control region around the $Z$-boson
mass peak, where the normalisations of the $Z$+light-jet and $Z$+heavy-flavour-jet
are simultaneously extracted with a combined fit and are found to be in
agreement with the theoretical expectation for $Z$+jets.
The uncertainty in the normalisation and in the flavour
composition of $W$+jets is assessed using data control regions. The total
normalisation and flavour fraction uncertainty is about 22\% for
$Wb(b)$ and $Wcc$, approximately 45\% for $Wc$, and about 23\% for
$W+$light-jets.
 
For the multijet background, a 30\% systematic uncertainty is assigned to the predicted yields, based on comparisons
with data yields in control regions similar to the signal region but
enriched in events from the multijet background; the $e$+jet
and $\mu$+jet events are treated as uncorrelated.
For the small diboson background, a 50\% normalisation uncertainty is
assigned and includes uncertainties in the inclusive cross-section and
additional jet production~\cite{ATL-PHYS-PUB-2016-002}.
 
\subsection{Detector response}
Uncertainties associated with leptons arise from the trigger, reconstruction,
identification, and isolation requirements, as well as the lepton momentum scale and
resolution. The reconstruction, identification and isolation efficiency for
electrons and muons, as well as the efficiency of the trigger, differ
slightly between data and simulation and are compensated for by
dedicated SFs.
Efficiency SFs are derived using  data and simulated samples of $Z\to \ell^+\ell^-$
($\ell=e,\mu$), and are applied to the simulation to
correct for differences. The effect of uncertainties in these SFs is propagated through the analysis.
The total uncertainty in efficiency SFs, for the high-\pt{} leptons,
is $<0.5\%$ for muons across the entire \pt
spectrum \cite{PERF-2015-10} and for electrons with $\pt>30~\GeV$, while it
exceeds $1\%$ for lower-\pt electrons~\cite{PERF-2017-03}. The corresponding uncertainties are at the level of few per-mill for the momentum scale both for electrons and muons, and of 5-10\% (4-7\%) for the electron (muon) momentum resolution, strongly depending
on the detector region and on the transverse momentum of the lepton.
Additional sources of uncertainty originate from the uncertainty in the corrections applied to
adjust the lepton momentum scale and resolution in the simulation to match those
in data. They are measured using reconstructed $Z\to \ell^+\ell^-$
and $J/\psi \to \ell^+\ell^-$ dilepton invariant mass distributions~\cite{PERF-2017-01,PERF-2015-10}.
To evaluate the effect of momentum scale and resolution uncertainties, the analysis is
repeated with the lepton momentum varied by $\pm 1\sigma$ and with the lepton momentum smeared, respectively.
A systematic uncertainty due to the electron charge misidentification is taken from Ref.~\cite{PERF-2017-01}.
Scale factors correcting for the differences in electron charge
misidentification rates between data and simulation are computed using $Z \to e^+ e^-$ events.
 
Uncertainties associated with jets arise from the efficiency of jet
reconstruction and identification based on the JVT variable, as well as the JES and the jet energy resolution (JER). Although the
observable \mlmu{} does not involve jets, the various jet
uncertainties impact the analysis through the event selection.
The JES and its uncertainty were derived by combining information
from test-beam data, LHC collision data and simulation~\cite{PERF-2016-04}.
The JES uncertainty is about 5.5\% for jets with $\pt =25~\gev$ and
decreases quickly with increasing jet $\pt$. It is below 1.5\% for central jets with
$\pt$ in the range of approximately $100~\gev$ to $1.5~\tev$.
The highest-\pt{} jet in this analysis has an average \pt{} of around 130~\GeV{}, with a typical range between 50~\GeV{} and 450~\GeV{}.
The uncertainty from the soft muon jet $p_\textrm{T}$ calibration affects
the measured top-quark mass marginally, through the event selection.
The magnitude of the JER uncertainty variation
is parameterised in jet $\pt$ and $\eta$~\cite{ATL-PHYS-PUB-2015-015}, and
the uncertainty is propagated by smearing the jet $\pt$ in the simulation. The uncertainty in the efficiency to
pass the JVT requirement is evaluated by varying the scale factors within their uncertainties~\cite{PERF-2014-03}.
 
The efficiencies of DJ tagging in simulated samples are corrected to match efficiencies
in data. Correction scale factors are derived for jets originating from
$b$-, $c$- and light-quarks separately in dedicated calibration analyses~\cite{PERF-2016-05,ATLAS-CONF-2018-001,ATLAS-CONF-2018-006}. For jets
originating from $b$- and $c$-quarks, SFs are derived as a function of \pt,
whereas the light-jet efficiency is scaled by $\pt$- and $\eta$-dependent
factors. Uncertainties in the correction scale factors are estimated by varying each source of
uncertainty up and down by one standard deviation and are taken as uncorrelated between $b$-jets,
$c$-jets, and light-jets. The same SFs are verified to also be applicable to jets containing soft muons by repeating the calibration procedure on a dedicated sample of events.
An additional set of MC-to-MC correction factors are introduced to
account for the different parton shower and hadronisation model used in the calibration samples
compared to those used in this analysis.
Furthermore, the efficiency of tagging the hadronic decays of
$\tau$-leptons in simulation is treated in the same way as the efficiency of
tagging $c$-jets, and a specific uncertainty is considered for this
simplified approach.
An additional check is performed
by changing the event selection such that there is always a DJ tagged
jet other than the SMT tagged jet in the event, and the measured top-quark mass
is consistent with the value measured using the nominal event selection.
 
The $\met$ reconstruction is affected by uncertainties associated with lepton
and jet energy scales and resolutions, which are appropriately
propagated to the $\met$ calculation.
Additional small uncertainties associated with the modelling of the underlying
event, in particular its impact on the $\pt$ scale and resolution of unclustered
energy, are also taken into account~\cite{PERF-2016-07}.
 
\begin{figure}
\centering
\includegraphics[width=0.48\textwidth]{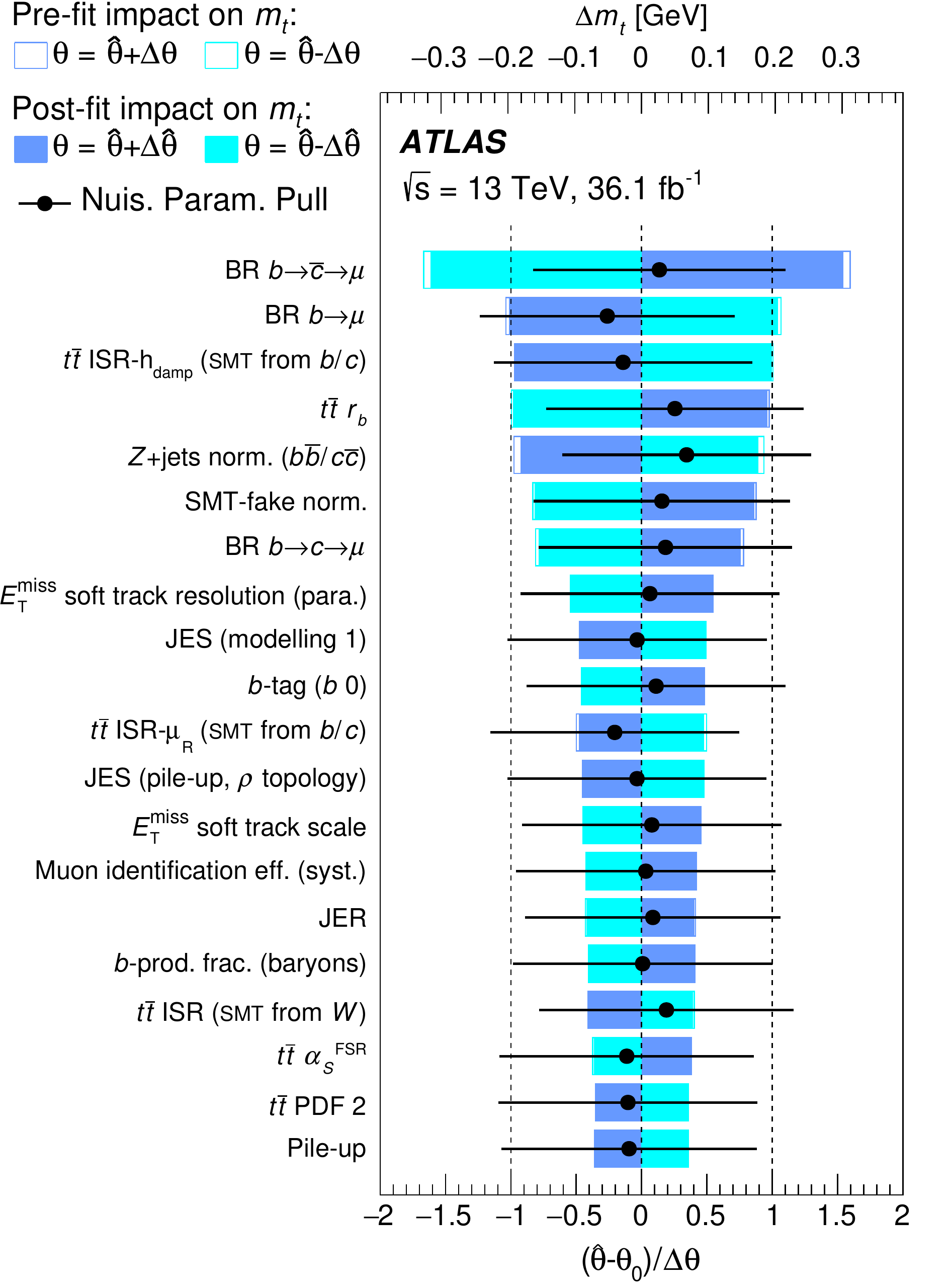}
\caption{Ranking, from top to bottom, of the main systematic
uncertainties (excluding recoil) showing the pulls
and the impact of the systematic uncertainties on $m_t$, from the
combined OS and SS binned-template profile likelihood fit to
data. The $\theta$'s represent each systematic variation.
The upper scale shows the uncertainty contribution to the measured top-quark mass. The PDF 2 is the set number 2 of the \textsc{PDF4LHC15} error set~\cite{PDF4LHC15}.}
\label{fig:fit_pulls_b}
\end{figure}

The total systematic uncertainty is computed from the variance $(\sigma^2)$ difference between the total uncertainty returned by
the fit and the data statistics uncertainty. The plot in Figure~\ref{fig:fit_pulls_b} shows the ranking of the main systematic uncertainties, including the pulls and the impact of constraining the systematic
uncertainties. Uncertainties in Figure~\ref{fig:fit_pulls_b} are from the individual nuisance
parameters. They may be included within a grouped category in
Table~\ref{tab:fit_systs}. The leading uncertainties are
due to the modelling of the $b$-quark fragmentation $b$-hadron and
decay. In particular, the BRs for direct and sequential decays are important because the
SMT muon $\pt$ is softer when it is produced from $c$-hadrons in a
cascade $b$-hadron decay, than when it comes directly from a
semileptonic $b$-hadron decay. The BRs also impact the charge-signs combination
of the primary lepton and the soft muon.
Nearly all of the main systematic uncertainties are
largely uncorrelated with those dominant in previous top-quark direct reconstruction measurements~\cite{TOPQ-2017-03}, and
the uncertainty from jet energy calibration is sub-dominant with a
value of $\pm0.13$~\GeV.


\FloatBarrier
\section{Conclusions}
\label{sec:conclusion}

 
A direct measurement of the top-quark mass has been performed using a technique that exploits a partial, leptonic-only,
invariant mass reconstruction of the top-quark decay products.
The analysis uses data corresponding to an integrated luminosity of
36.1~fb$^{-1}$ of $\sqrt{s} = 13$~\TeV\ $pp$ collisions provided by the Large Hadron Collider and recorded by the ATLAS detector, and is
based on  the invariant mass, $m_{\ell\mu}$, of the lepton $\ell$ (with $\ell=e,\mu$) from the $W$-boson decay and the muon
$\mu$ from a semileptonic decay of a $b$-hadron. A binned-template profile likelihood fit to the $m_{\ell\mu}$ distribution is performed to determine
the most probable top-quark mass value.
The result, $m_{t} = 174.41\pm0.39~(\textrm{stat.})\pm0.66~(\textrm{syst.})\pm0.25~(\textrm{recoil})~\textrm{GeV}$, 
corresponds to the most precise single measurement to date of the top-quark mass from the direct reconstruction of its decay products
by the ATLAS Collaboration, and more precise than those performed previously with similar techniques~\cite{Aaltonen:2009zi,CMS-TOP-15-014}.
The third uncertainty arises from using a recently developed setup of the \textsc{Pythia8} parton shower gluon-recoil scheme in top quark decays.
 
Taking into account the correlation between uncertainties, the result is consistent at the level of approximately two standard deviations with the current ATLAS combination of top-quark mass measurements
from the reconstruction of the top-quark decay~\cite{TOPQ-2017-03}. A similar level of consistency is found with the equivalent combination at CMS~\cite{CMS-TOP-14-022}, while agreement with the latest Tevatron combination~\cite{TEVEWWG} is good.
Agreement within one standard deviation is also found with the indirect
prediction of the top-quark mass from global electroweak fits~\cite{EWfits}. The main sources of systematic uncertainty are due to the modelling of
top-quark pair production and $b$-quark fragmentation and decay, with uncertainties from backgrounds also being significant. On the other hand, the uncertainty due to the calibration of the jet energies is sub-dominant, which is advantageous in future combinations of this result with those from the standard reconstruction of the top-quark decay products.


\section*{Acknowledgements}


We thank CERN for the very successful operation of the LHC, as well as the
support staff from our institutions without whom ATLAS could not be
operated efficiently.
 
We acknowledge the support of
ANPCyT, Argentina;
YerPhI, Armenia;
ARC, Australia;
BMWFW and FWF, Austria;
ANAS, Azerbaijan;
CNPq and FAPESP, Brazil;
NSERC, NRC and CFI, Canada;
CERN;
ANID, Chile;
CAS, MOST and NSFC, China;
Minciencias, Colombia;
MEYS CR, Czech Republic;
DNRF and DNSRC, Denmark;
IN2P3-CNRS and CEA-DRF/IRFU, France;
SRNSFG, Georgia;
BMBF, HGF and MPG, Germany;
GSRI, Greece;
RGC and Hong Kong SAR, China;
ISF and Benoziyo Center, Israel;
INFN, Italy;
MEXT and JSPS, Japan;
CNRST, Morocco;
NWO, Netherlands;
RCN, Norway;
MEiN, Poland;
FCT, Portugal;
MNE/IFA, Romania;
MESTD, Serbia;
MSSR, Slovakia;
ARRS and MIZ\v{S}, Slovenia;
DSI/NRF, South Africa;
MICINN, Spain;
SRC and Wallenberg Foundation, Sweden;
SERI, SNSF and Cantons of Bern and Geneva, Switzerland;
MOST, Taiwan;
TENMAK, T\"urkiye;
STFC, United Kingdom;
DOE and NSF, United States of America.
In addition, individual groups and members have received support from
BCKDF, CANARIE, Compute Canada and CRC, Canada;
PRIMUS 21/SCI/017 and UNCE SCI/013, Czech Republic;
COST, ERC, ERDF, Horizon 2020 and Marie Sk{\l}odowska-Curie Actions, European Union;
Investissements d'Avenir Labex, Investissements d'Avenir Idex and ANR, France;
DFG and AvH Foundation, Germany;
Herakleitos, Thales and Aristeia programmes co-financed by EU-ESF and the Greek NSRF, Greece;
BSF-NSF and MINERVA, Israel;
Norwegian Financial Mechanism 2014-2021, Norway;
NCN and NAWA, Poland;
La Caixa Banking Foundation, CERCA Programme Generalitat de Catalunya and PROMETEO and GenT Programmes Generalitat Valenciana, Spain;
G\"{o}ran Gustafssons Stiftelse, Sweden;
The Royal Society and Leverhulme Trust, United Kingdom.
 
The crucial computing support from all WLCG partners is acknowledged gratefully, in particular from CERN, the ATLAS Tier-1 facilities at TRIUMF (Canada), NDGF (Denmark, Norway, Sweden), CC-IN2P3 (France), KIT/GridKA (Germany), INFN-CNAF (Italy), NL-T1 (Netherlands), PIC (Spain), ASGC (Taiwan), RAL (UK) and BNL (USA), the Tier-2 facilities worldwide and large non-WLCG resource providers. Major contributors of computing resources are listed in Ref.~\cite{ATL-SOFT-PUB-2021-003}.


\printbibliography
 
\clearpage
 
\begin{flushleft}
\hypersetup{urlcolor=black}
{\Large The ATLAS Collaboration}

\bigskip

\AtlasOrcid[0000-0002-6665-4934]{G.~Aad}$^\textrm{\scriptsize 103}$,
\AtlasOrcid[0000-0002-5888-2734]{B.~Abbott}$^\textrm{\scriptsize 122}$,
\AtlasOrcid[0000-0002-7248-3203]{D.C.~Abbott}$^\textrm{\scriptsize 104}$,
\AtlasOrcid[0000-0002-2788-3822]{A.~Abed~Abud}$^\textrm{\scriptsize 36}$,
\AtlasOrcid[0000-0002-1002-1652]{K.~Abeling}$^\textrm{\scriptsize 55}$,
\AtlasOrcid[0000-0002-2987-4006]{D.K.~Abhayasinghe}$^\textrm{\scriptsize 95}$,
\AtlasOrcid[0000-0002-8496-9294]{S.H.~Abidi}$^\textrm{\scriptsize 159}$,
\AtlasOrcid[0000-0002-8279-9324]{O.S.~AbouZeid}$^\textrm{\scriptsize 42}$,
\AtlasOrcid{N.L.~Abraham}$^\textrm{\scriptsize 149}$,
\AtlasOrcid[0000-0001-5329-6640]{H.~Abramowicz}$^\textrm{\scriptsize 154}$
% ...
\bigskip
\\

% ...
$^{36}$CERN, Geneva; Switzerland.\\
% ...
$^{42}$Niels Bohr Institute, University of Copenhagen, Copenhagen; Denmark.\\
% ...
$^{55}$II. Physikalisches Institut, Georg-August-Universit\"{a}t G\"ottingen, G\"ottingen; Germany.\\
% ...
$^{95}$Department of Physics, Royal Holloway University of London, Egham; United Kingdom.\\
% ...
$^{103}$CPPM, Aix-Marseille Universit\'e, CNRS/IN2P3, Marseille; France.\\
$^{104}$Department of Physics, University of Massachusetts, Amherst MA; United States of America.\\
% ...
$^{122}$Homer L. Dodge Department of Physics and Astronomy, University of Oklahoma, Norman OK; United States of America.\\
% ...
$^{149}$Department of Physics and Astronomy, University of Sussex, Brighton; United Kingdom.\\
% ...
$^{154}$Raymond and Beverly Sackler School of Physics and Astronomy, Tel Aviv University, Tel Aviv; Israel.\\
% ...
$^{159}$Department of Physics, University of Toronto, Toronto ON; Canada.\\
% ...

\end{flushleft}


\clearpage

\end{document}